\documentclass[aps,prd,nofootinbib,floatfix,superscriptaddress,secnumarabic]{revtex4}
\usepackage{bm}
\usepackage{graphicx}                                                                                                                                                                                    
\usepackage{epsf}
\usepackage{epsfig}

\usepackage{amsmath}
\usepackage{amsfonts,amssymb}

\usepackage{colordvi}
\usepackage{color}
\usepackage{lscape}
\usepackage{rotfloat}
\usepackage{rotating}

\usepackage{dsfont}
\usepackage{slashed}
\usepackage{booktabs}

\newcommand{\be}{\begin{equation}}
\newcommand{\ee}{\end{equation}}
\newcommand{\bea}{\begin{eqnarray}}
\newcommand{\eea}{\end{eqnarray}}

\newcommand{\MSbar}{{\overline{\rm MS}}}

\newcommand{\gtilde}{\frac{g^2}{16 \, \pi^2}\; }
\newcommand{\qslash}{{\not{\hspace{-0.001cm}q}}}

\newcommand{\qqq}{\displaystyle{\not}q_3}
\newcommand{\qq}{\displaystyle{\not}q_2}
\newcommand{\bpsi}{\overline{\psi}}
\newcommand{\Dr}{\displaystyle{\not}{\overrightarrow{D}}}
\newcommand{\Dl}{\displaystyle{\not}{\overleftarrow{D}}}
\newcommand{\partl}{\displaystyle{\not}{\overleftarrow{\partial}}}
\newcommand{\partr}{\displaystyle{\not}{\overrightarrow{\partial}}}
\newcommand{\DDr}{\overrightarrow{D}}
\newcommand{\DDl}{\overleftarrow{D}}

\begin{document}


\vspace{1cm}

\centerline{\huge The chromomagnetic operator on the lattice}

\vspace{1cm}

\centerline{\Large M.~Constantinou$^{(a)}$, M.~Costa$^{(a)}$, R.~Frezzotti$^{(b)}$, V.~Lubicz$^{(c,d)}$}

\centerline{\Large G.~Martinelli$^{(e,f)}$, D.~Meloni,$^{(c,d)}$ H.~Panagopoulos$^{(a)}$, S.~Simula$^{(d)}$}

\vspace{0.5cm}

\centerline{\it $^{(a)}$ Department of Physics, University of Cyprus, Nicosia, CY-1678, Cyprus}

\centerline{\it $^{(b)}$ Dip. di Fisica, Universit\`a di Roma ``Tor Vergata'' and INFN, Sezione di ``Tor Vergata'',}
\centerline{\it Via della Ricerca Scientifica 1, I-00133 Rome, Italy}

\centerline{\it $^{(c)}$ Dip. di Fisica, Universit\`a  Roma Tre, Via della Vasca Navale 84, I-00146 Rome, Italy}

\centerline{\it $^{(d)}$ INFN, Sezione di Roma Tre, Via della Vasca Navale 84, I-00146 Rome, Italy}

\centerline{\it $^{(e)}$ SISSA, Via Bonomea 265, I-34136, Trieste, Italy}

\centerline{\it $^{(f)}$ Dip. di Fisica, Universit\`a  di Roma ``La Sapienza" and INFN, Sezione di Roma,}
\centerline{\it P.le Aldo Moro 2, I-00185 Rome, Italy} 

\vspace{0.5cm}

\begin{figure}[!htb]
\centerline{\includegraphics{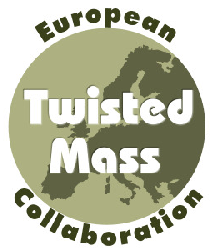}}
\end{figure}

\vspace{0.5cm}

\centerline{\bf Abstract}

\vspace{0.25cm}

We present our study of the renormalization of the chromomagnetic operator, ${\cal O}_{CM}$, which appears in the effective Hamiltonian describing $\Delta S = 1$ transitions in and beyond the Standard Model.

We have computed, perturbatively to one-loop, the relevant Green's functions with two (quark-quark) and three (quark-quark-gluon) external fields, at nonzero quark masses, using both the lattice and dimensional regularizations.
The perturbative computation on the lattice is carried out using the maximally twisted-mass action for the fermions, while for the gluons we employed the Symanzik improved gauge action for different sets of values of the Symanzik coefficients.
We have identified all the operators which can possibly mix with ${\cal O}_{CM}$, including lower dimensional and non gauge invariant operators, and we have calculated those elements of the mixing matrix which are relevant for the renormalization of ${\cal O}_{CM}$.

We have also performed numerical lattice calculations to determine non-perturbatively the mixings of the chromomagnetic operator with lower dimensional operators, through proper renormalization conditions.
For the first time the $1 / a^2$-divergent mixing of the chromomagnetic operator with the scalar density has been determined non-perturbatively with high precision.  Moreover, the $1 / a$-divergent mixing with the pseudoscalar density, due to the breaking of parity within the twisted-mass regularization of QCD, has been calculated non-perturbatively and found to be smaller than its one-loop perturbative estimate.
The QCD simulations have been carried out using the gauge configurations produced by the European Twisted Mass Collaboration with $N_f = 2 + 1 + 1$ dynamical quarks, which include in the sea, besides two light mass degenerate quarks, also the strange and charm quarks with masses close to their physical values. 

\newpage

\pagestyle{plain}

\section{Introduction}
\label{Intro}

A very natural explanation for the extraordinary success of the Standard Model (SM) in describing the electro-weak and strong interactions at the fundamental level is that the SM Lagrangian contains all relevant operators of dimension $d \le 4$ composed by the (already observed) elementary particle fields and compatible with the principles of Lorentz invariance and gauge symmetry. The effects of higher dimension ($d > 4$) effective operators, which are not included in the SM Lagrangian, are expected to be naturally small, being suppressed by negative powers of the high energy scale $M$ characterising the physics beyond the SM, as $M^{4-d}$ (up to logarithmic corrections).

In this picture, a special role is played by the operators of dimension $d = 5$, as their contribution is suppressed by only one power of the high scale. In the leptonic sector, an important example of $d = 5$ operator is represented by the Weinberg operator for neutrino masses~\cite{Weinberg:1980bf}, composed by two lepton doublets and two Higgs fields. After the occurrence of spontaneous electroweak symmetry breaking this operator leads to a natural explanation of the small light-neutrino masses, which are thus predicted to be inversely proportional to the large scale $M$. 

In the quark sector the $d = 5$ magnetic operators, which induce $\Delta F = 1$ flavor changing transitions, are of relevant phenomenological interest. In the strangeness changing $\Delta S = 1$ case, for instance, these magnetic operators contribute to both CP conserving and CP violating rare kaon decays, as well as to $K^0 - \bar{K}^0$ mixing and to the CP violating ratio $\varepsilon^\prime / \varepsilon$. In a large class of new physics models these contributions can be substantially larger than in the SM, which motivates the interest in studying their effects. This is the case, for instance, of generic supersymmetric extensions of the SM, in which $\Delta S=1$ transitions described by the magnetic operators are mediated by the strong interactions through virtual gluino exchanges~\cite{Buras:1999da,D'Ambrosio:1999jh}.

For definiteness, let us consider in the following $\Delta S=1$ transitions. In both the SM and beyond, the low-energy effective Hamiltonian contains four magnetic operators of dimension 5,
\be
H^{\Delta S = 1,\ d=5}_{\rm eff} = \sum_{i=\pm} (C^i_\gamma Q^i_\gamma + C^i_g Q^i_g) + {\rm h.c.}
\ee
which are defined as:
\bea
\label{Qgammapm}
&& Q_\gamma^\pm = \frac{Q_d\,e}{16 \pi^2} \left(\bar \psi_{sL} \,\sigma^{\mu\nu}\, F_{\mu\nu} \, \psi_{dR} 
\pm \bar \psi_{sR} \,\sigma^{\mu\nu}\, F_{\mu\nu} \, \psi_{dL}  \right) \ , \\
\label{Qgpm}
&& Q_g^\pm = \frac{g}{16 \pi^2} \left(\bar \psi_{sL} \,\sigma^{\mu\nu}\, G_{\mu\nu} \, \psi_{dR}   
\pm \bar \psi_{sR} \,\sigma^{\mu\nu}\, G_{\mu\nu} \, \psi_{dL} \right) \ .
\eea

In the above expressions, $F_{\mu\nu}$ and $G_{\mu\nu}$ represent  the electromagnetic and strong field strength tensors respectively, $\psi_s$ and $\psi_d$ are the strange and down quark fields and the subscripts $R,L$ denote the left/right chiral structure $(1 \pm \gamma^5)$. The coefficients $C^i_{\gamma}$ and $C^i_{g}$, multiplying the electromagnetic (EMO) and chromomagnetic (CMO) operators in the effective Hamiltonian, contain the effects of the physics at short distance and they depend on the specific structure of the new physics model. These coefficients can be calculated perturbatively via the OPE. The long distance effects of the strong interactions are encoded in the operator matrix elements and thus require, for their evaluation, a non-perturbative method, primarily a lattice QCD calculation. 

The matrix element of the EMO between kaon and pion states may be relevant in the CP-violating part of the $K \to \pi \ell^+ \ell^-$ semileptonic decays (see Ref.~\cite{FlaviaNet}) and its determination offers, for instance, the possibility to put bounds on the supersymmetric couplings related to the splitting of the off-diagonal entries in the down-type squark mass matrix. The matrix element $\langle \pi | Q_\gamma^+ | K \rangle$ has been computed on the lattice both in the quenched approximation \cite{Becirevic:2000zi} and with $N_f = 2$ flavors of degenerate sea quarks \cite{Baum:2011rm}.

Several matrix elements of the CMO between kaon and pion states are of phenomenological interest for supersymmetric extensions of the SM. 
The matrix element $\langle \pi^0 | Q_g^+ | K^0 \rangle$ may provide contributions to the $K^0 - \bar{K}^0$ mixing amplitude (see Ref.~\cite{D'Ambrosio:1999jh}), while the matrix element $\langle \pi^+ \pi^- | Q_g^- | K^0 \rangle$ may play a role for determining $\varepsilon^\prime / \varepsilon$ and for the $\Delta I = 1/2$ rule (see Ref.~\cite{Buras:1999da}). Finally the matrix element $\langle \pi^+ \pi^+ \pi^- | Q_g^+ | K^+ \rangle$ may contribute to the CP-violating part of the $K \to \pi \pi \pi$ decays \cite{D'Ambrosio:1999jh}.
All the above-mentioned matrix elements can be parameterized in terms of suitably defined B-parameters:
\bea
    \label{ME}
    \langle \pi^0 | Q_g^+ | K^0 \rangle & = & - \frac{1}{\sqrt{2}} \frac{11}{32\pi^2}\, \frac{M_K^2 \, (p_K \cdot p_\pi)}{m_s + m_d} \, B_{CMO}^{K\pi} \ , \nonumber \\
    \langle \pi^+ \pi^- | Q_g^- | K^0 \rangle & = & i\, \frac{11}{32\pi^2}\, \frac{M_K^2 \, M_\pi^2}{f_\pi\,(m_s + m_d)} \, B_{CMO}^{K2\pi} \ , \\
    \langle \pi^+ \pi^+ \pi^- | Q_g^+ | K^+ \rangle & = & - \frac{11}{16\pi^2}\, \frac{M_K^2 \, M_\pi^2}{f_\pi^2\,(m_s + m_d)} \, B_{CMO}^{K3\pi} \ .\nonumber
\eea

At leading order (LO) in Chiral Perturbation Theory (ChPT) the CMO has a single representation in terms of pseudo-Goldstone boson fields~\cite{Bertolini:1994qk}.
Therefore, the three B-parameters appearing in Eq.~(\ref{ME}) are related by chiral symmetry, which predicts at LO their equality: 
 \be
    B_{CMO}^{K\pi} = B_{CMO}^{K2\pi} = B_{CMO}^{K3\pi} = B_{CMO} ~ .
    \label{BCMO}
 \ee

A lattice calculation of the matrix elements of the CMO is challenging, particularly when more than one pion are considered in the final state.
Even in the case of only one final pion, which corresponds to the matrix element $\langle \pi | Q_g^+ | K \rangle$ of the operator $Q_g^+ = (g / 16 \pi^2) ~ \bpsi_s\, \sigma_{\mu \nu}\, G_{\mu \nu} \psi_d$, no results have been produced so far.
The main difficulty, with respect to the EMO case, is that strong interactions induce a mixing of the CMO with operators of lower dimension, with coefficients which are power divergent with the cutoff, which on the lattice is the inverse of the lattice spacing $1/a$.
The leading divergence of the bare CMO, which is of order $1/a^2$, is determined by the mixing with the dimension-3 scalar operator $\bar \psi_s \psi_d$. 
Its coefficient must be evaluated in a fully non-perturbative way, since non-perturbative effects, e.g., factors of the form $a \Lambda_{QCD}$, combined with factors which diverge as inverse powers of the lattice spacing can give finite (or even divergent) contributions \cite{Maiani:1992vl}. 

In order to define the properly renormalized CMO, besides the subtraction of the lower dimension operators, the mixing with equal dimension ($d = 5$) operators, including the CMO itself, must also be taken into account.
This mixing is only logarithmically divergent and can be thus evaluated in perturbation theory. 
The one-loop calculation of the corresponding renormalization factor and mixing coefficients is one aim of the present study. 
Specifically, we have considered a lattice regularization of QCD defined by a generic class of Symanzik improved gluon actions and a twisted-mass quark action \cite{Frezzotti:2000nk,Frezzotti:2003ni}. 
By investigating the symmetry properties of this action, we have shown that the renormalized CMO mixes with a total of 13 operators (including itself), of which seven are not present on-shell; among them, there will be non-gauge invariant (but BRST invariant) operators as well.
For on-shell matrix elements, the mixing assumes the general form:
\bea
g_0\,\bpsi_s\, \sigma_{\mu \nu}\, G_{\mu \nu} \psi_d  &=& Z_{1}\, \left[ g_0\,\bpsi_s\, \sigma_{\mu \nu}\, G_{\mu \nu} \psi_d \right]^R +  Z_{2}\,\left[ (m_{d}^2+m_{s}^2)\bpsi_s\psi_d\right]^R  +  Z_{3}\,\left[m_{d}\,m_{s}\bpsi_s\psi_d\right]^R \nonumber \\
&&+ Z_{4}\, \left[ \Box \left( \bpsi_s \psi_d \right) \right]^R +  Z_{12} \left[\,i\,(r_d\,m_{d}+r_s\,m_{s})\bpsi_s\gamma_5\psi_d\right]^R + Z_{13}\left[\bpsi_s\,\psi_d\right]^R\,, 
\label{eq:CMO_mixing}
\eea
where $R$ denotes the corresponding renormalized operators, $Z_{12} \propto 1/a$, $Z_{13} \propto 1/a^2$, $r_{s(d)}$ is the Wilson hopping parameter of the strange (down) quark, and we have evaluated the renormalization factor $Z_1$ and the mixing coefficients $Z_2 - Z_{13}$ at one loop.
Note that the presence of the mixing with the pseudoscalar density is due to the parity violation in the twisted-mass formulation of QCD on the lattice.
As discussed above, the power-divergent coefficients $Z_{12}$ and $Z_{13}$ require an independent non-perturbative determination.

A strategy to implement non-perturbatively the subtraction of the mixings of the CMO with the lower dimension operators has been anticipated in Refs.~\cite{Constantinou:2013zqa,Constantinou:2014cra, Constantinou:2014wna}.
In this work we apply it for obtaining the first non-perturbative determinations of the power-divergent mixings $Z_{13}$ and $Z_{12}$.
To this end we have used the gauge configurations produced at three values of the lattice spacing (between $\simeq 0.6$ and $\simeq 0.9$ fm) by the European Twisted Mass Collaboration (ETMC) with $N_f = 2 + 1 + 1$ dynamical quarks, which include in the sea, besides two light mass degenerate quarks, also the strange and charm quarks with masses close to their physical values \cite{Baron:2010bv,Baron:2010th,Baron:2011sf}.
It turns out that the one-loop perturbative estimate of $Z_{13}$ differs only by less than $10 \%$ from the non-perturbative results at the three values of the lattice spacing, while the non-perturbative determination of $Z_{12}$ is found to be smaller than the corresponding one-loop perturbative result. 
This finding suggests that, together with our non-perturbative determinations of the mixing coefficients $Z_{13}$ and $Z_{12}$, the perturbative estimates of the renormalization factor $Z_1$ and of the mixing coefficients $Z_2$, $Z_3$ and $Z_4$ may be used for the determination of the (renormalized) CMO matrix element. Preliminary results for such a matrix element between pion and kaon states have been presented in Ref.~\cite{Constantinou:2014tea} and the final ones will be the subject of a forthcoming publication~\cite{Constantinou:future}.

The outline of this paper is as follows.
Section~\ref{sec1} provides a brief theoretical background in which we introduce the symmetries of the employed actions and the transformation properties of all candidate operators which can mix with ${\cal O}_{CM}$ at the quantum level. 
Section~\ref{RFs} contains a summary of the computational procedure for the Green's functions of the chromomagnetic operator.
This Section is divided in two subsections. In Subsection~\ref{sec2.1}, calculating the 2-point and 3-point Green's function of ${\cal O}_{CM}$ in dimensional regularization (DR), we construct a set of eleven independent equations for the disentanglement of the mixing coefficients.
We present these coefficients in the $\MSbar$ renormalization scheme.
On the other hand in Subsection~\ref{sec2} using the lattice formulation and the results which we found in Subsection~\ref{sec2.1}, we calculate the mixing coefficients on the lattice, again in the $\MSbar$ renormalization scheme.
In Section~\ref{Nonpert} we describe the first non-perturbative, high-precision determination of the $1 / a^2$-divergent mixing of the chromomagnetic operator with the scalar density, using the ETMC gauge configurations with $N_f = 2 +1 + 1$ produced at three values of the lattice spacing.
We also describe the first non-perturbative calculation of the $1 / a$-divergent mixing of the chromomagnetic operator with the pseudoscalar density using the renormalization condition introduced in Ref.~\cite{Constantinou:2013zqa}.
Finally, we conclude in Section~\ref{summary} with a discussion of our results and possible future extensions of our work.
For completeness, we have included two Appendices containing the mixing coefficients $Z_i$ (Appendix~\ref{app:mixCoef}) and the one-loop perturbative renormalization factors $Z_c$, $Z_\psi$, $Z_m$, $Z_A$ and $Z_g$ on the lattice (Appendix~\ref{app:Z}).
Preliminary results for the above coefficients have been already presented in Ref.~\cite{Constantinou:2013zqa,Constantinou:2014cra, Constantinou:2014wna,Constantinou:2014tea}.

\section{Symmetries of the Action and Transformation Properties of Operators}
\label{sec1}

We start by studying the mixing of the chromomagnetic operator\footnote{In our notation $g_0$ is the bare coupling constant, $\psi_{s,d}$ are the s- and d-quark fields, $G_{\mu \nu}$ is the gluon tensor and $\sigma_{\mu \nu} \equiv (i/2) [\gamma_\mu,\gamma_\nu]$.}:
\be
{\cal O}_{CM} = g_0\,\bpsi_s\, \sigma_{\mu \nu}\, G_{\mu \nu} \psi_d,  
\label{eq:OCM}
\ee
using both DR and lattice regularization (L). 
On the lattice we use the fermion setup studied in Refs.~\cite{Frezzotti:2003ni, Frezzotti:2000nk}; in particular, valence quarks are described by the twisted-mass/Osterwalder-Seiler (OS) action at maximal twist, which in the physical basis reads: 
\be
S_F[\psi_f,\bar \psi_f,U]= a^4 \sum_f\,\sum_x\,\bar \psi_f(x) \Big{[}\gamma\cdot\widetilde\nabla
-i\gamma_5 W_{\rm{cr}}(r_f)+m_f\Big{]}\psi_f(x)\,,
\label{Action1}
\ee
where
\bea
\gamma\cdot\widetilde\nabla&\equiv&\frac{1}{2}\sum_\mu\gamma_\mu
(\nabla^\star_\mu+\nabla_\mu)\, ,\\
W_{\rm{cr}}(r_f)&\equiv&-a\frac{r_f}{2}\sum_\mu\nabla^\star_\mu\nabla_\mu+
M_{\rm{cr}}(r_f)\, ,
\eea
$r_f$ is the Wilson parameter for the flavor $f=u,\,d,\,s$ and
$M_{\rm{cr}}(r_f)$ is  the corresponding critical quark mass.

The full fermion action includes also a part describing sea quarks, and possibly a ghost part (to compensate the valence quark determinant for the partially quenched flavors)~\cite{Frezzotti:2004wz}; these parts will not be needed in our perturbative calculation. 
For the gluon part we employ the Symanzik improved action:
\bea
\hspace{-1cm}
S_G=\frac{2}{g_0^2} \Bigl[ &c_0& \sum_{\rm plaq.} {\rm
    Re\,Tr\,}\{1-U_{\rm plaq.}\}
\,+\, c_1 \sum_{\rm rect.} {\rm Re \, Tr\,}\{1- U_{\rm rect.}\}
\nonumber \\
+ &c_2& \sum_{\rm chair} {\rm Re\, Tr\,}\{1-U_{\rm chair}\}
\,+\, c_3 \sum_{\rm paral.} {\rm Re \,Tr\,}\{1-U_{\rm
  paral.}\}\Bigr]\,,
\label{Symanzik}
\eea
where the Wilson loops are products of consecutive links in the directions $(\mu, \nu, -\mu, -\nu)$, $(\mu, \nu, \nu, -\mu, -\nu, -\nu)$, $(\mu, \nu, -\mu, \rho, -\nu, -\rho)$, $(\mu, \nu, \rho, -\mu, -\nu, -\rho)$ for $U_{\rm plaq.}$,$U_{\rm rect.}$,$U_{\rm chair}$ and $U_{\rm paral.}$, respectively.
The Symanzik coefficients $c_0,c_1,c_2,c_3$ may take arbitrary values, subject to the constraint: 
\be
c_0 + 8 c_1 + 16 c_2 + 8 c_3 = 1\,,
\label{norm}
\ee
which ensures the correct classical continuum limit. Our results
(Section~\ref{sec2}) will be provided for some of the most popular
choices for $c_i$\,.

There exist certain symmetries of the action (valid both in the
continuum and lattice formulation of the theory) which reduce
considerably the number of operators that can possibly mix with ${\cal
O}_{CM}$ at the quantum level. These symmetries are defined by means of the 
discrete transformations ${\cal{P}}$ (continuum parity),
\be
{\cal{P}}:\left \{\begin{array}{ll}
&\hspace{-.3cm} U_0(x)\rightarrow U_0(x_{\cal{P}})\, ,\qquad
U_k(x)\rightarrow
U_k^{\dagger}(x_{\cal{P}}-a\hat{k})\, ,\qquad k=1,2,3\\
&\hspace{-.3cm} \psi_f(x)\rightarrow \gamma_0  \psi_f(x_{\cal{P}})\\
&\hspace{-.3cm}\bar{ \psi}_f(x)
\rightarrow\bar{ \psi}_f(x_{\cal{P}})\gamma_0\,,
\end{array}\right . \label{PAROP}
\ee
where $x_{\cal{P}}=(-{\bf{x}},x_0)$ and $\hat\mu$ is the unit vector in the $\mu$-direction,
\be
\hspace{-7.cm}{\cal{D}}_d : \left \{\begin{array}{lll}
U_\mu(x)&\rightarrow U_\mu^\dagger(-x-a\hat\mu) \\
 \psi_f(x)&\rightarrow e^{3i\pi/2}  \psi_f(-x)  \\
\bar{ \psi}_f(x)&\rightarrow e^{3i\pi/2} \bar{ \psi}_f(-x),
\end{array}\right . \label{FieldTransform}
\ee
which, besides  inverting $ x \to -x$, counts the parity of the dimension $d$ of each operator by multiplying it by $e^{i \pi d}$,
\be
{\cal{R}}_5=\prod_f {\cal{R}}_{f\,5}\, , \qquad {\cal{R}}_{f\,5} :
\left
\{\begin{array}{ll}
 \psi_f&\rightarrow\gamma_5  \psi_f  \\
\bar{ \psi}_f&\rightarrow -\bar{ \psi}_f\gamma_5\,,
\end{array}\right .
\label{psibarR5}
\ee
${\mathcal {C}}$ (charge conjugation; $^{\,T}$ means transpose)
\be
{\mathcal {C}}:\left \{\begin{array}{ll}
&\hspace{-.3cm}\psi(x)\rightarrow i\gamma_0 \gamma_2 \bar{ \psi}(x)^{T}\\
&\hspace{-.3cm}\bar{\psi}(x)\rightarrow
-{\psi}(x)^{T}i\gamma_0\gamma_2\\
&\hspace{-.3cm}U_\mu(x)\rightarrow U_\mu^{\star}(x)\, ,\quad
\mu=0,1,2,3\,,
\end{array}\right . 
\label{Chargeconjugation}
\ee
and $\cal{S}$ (exchange between the s and the d quark)
\be
{\mathcal {S}}:\left
\{\begin{array}{ll}
 \psi_s(x)&\leftrightarrow  \psi_d(x)  \\
\bar{\psi}_s(x)&\leftrightarrow  \bar{\psi}_d(x)  \\
m_s&\leftrightarrow m_d\,.
\end{array}\right .
\label{exchangesd}
\ee

In terms of the above transformations, the symmetries  of the action are\footnote{Note
that, in the case of $r_s = -r_d$, $\cal{CPS}$ will not be a symmetry of
the ${\it{\rm valence}}$ part of the action which contains a u
quark, since it will require $r_u \to -r_u$. However, the u quark can be
dropped from the valence part of the action, since our operator does not
contain u quarks, and therefore the Green's functions of interest will also
not contain any external u quarks. Nonetheless, it is important to note that
the sea quark part of the action is symmetric even in the presence of u, since it 
is an even function of the Wilson coefficients $r_f$ (by virtue also of $M_{\rm{cr}}(-r_f)= -M_{\rm{cr}}(r_f)$)~\cite{Frezzotti:2004wz}.}:
\bea
&\bullet& {\cal{P}}\times {\cal D}_d \times (m \to - m){\rm,\, where\,} m\, {\rm are\, all\, masses\, except}\, M_{\rm{cr}} \nonumber\\
&\bullet& {\cal{D}}_d\times {\cal R}_5\nonumber\\
\label{symmetries}
&\bullet& {\cal{C}}\times{\cal{S}}, {\rm ~ if ~ r_s = r_d}\\\nonumber
&\bullet& {\cal{C}}\times{\cal{P}}\times{\cal{S}}, {\rm ~ if ~ r_s = -r_d}\,.\nonumber
\eea

In order to identify which operators can possibly mix with ${\cal O}_{CM}$,
we examine the transformation properties of all candidate operators
under the above symmetries; admissible operators must transform in the
same way as ${\cal O}_{CM}$. Furthermore, by general renormalization theorems, these
operators must be gauge invariant, or else they must vanish by the
equations of motion.

\begin{table}
\begin{center}
\begin{tabular}{| c|  p{7cm} | c| c| c | c|}
\hline
\multicolumn{2}{|l|}{Operators}  &
${\cal{P}}\times {\cal D}_d \times$ & ${\cal{D}}_d\times {\cal R}_5$ & ${\cal{C}}\times{\cal{S}}$ & 
${\cal{C}}\times{\cal{P}}\times{\cal{S}}$\\

\multicolumn{2}{|l|}{}  & $(m \to - m)$ & $\,$ & if\,\,$r_s=r_d$& if\,\,$r_s=-r_d$\\\hline

\multicolumn{3}{l}{Dimension 3 operators}\\[0.5ex] \hline

$\,\checkmark$& $\,\bpsi_s\psi_d$  & $-$ & + & + & +\\[0.5ex]\hline

$\,$& $\,i\,\bpsi_s\gamma_5\psi_d$  & + & + & + & $-$\\[0.5ex]\hline

\multicolumn{3}{l}{Dimension 4 operators}\\[0.5ex] \hline

$\,$& $\,(m_{d}+m_{s})\bpsi_s\psi_d$ & + & + & + & +\\[0.5ex]\hline

$\,$& $\,(m_{d}-m_{s})\bpsi_s\psi_d$ & + & + & $-$ & $-$\\[0.5ex]\hline

$\,(+)$& $\,i\,(m_{d}+m_{s})\bpsi_s\gamma_5\psi_d$ & $-$ & +
& + & $-$\\[0.5ex]\hline

$\,(-)$& $\,i\,(m_{d}-m_{s})\bpsi_s\gamma_5\psi_d$ & $-$ &
+ & $-$ & + \\[0.5ex]\hline

$\,$& $\,\bpsi_s(\Dr+m_d)\psi_d
+\bpsi_s(-\Dl +m_s)\psi_d$&
+ & + & + & + \\[0.5ex]\hline

$\,$& $\,\bpsi_s(\Dr+m_d)\psi_d
-\bpsi_s(-\Dl +m_s)\psi_d$
& + & + & $-$ & $-$\\[0.5ex]\hline

$\,(+)$& $\,i\,\bpsi_s\gamma_5(\Dr+m_d)\psi_d
  +i\,\bpsi_s(-\Dl +m_s)\gamma_5\psi_d$&
$-$ & + & + & $-$ \\[0.5ex]\hline

$\,(-)$&$\,i\,\bpsi_s\gamma_5(\Dr+m_d)\psi_d
  -i\,\bpsi_s(-\Dl +m_s)\gamma_5\psi_d$&
$-$ & + & $-$ & + \\[0.5ex]\hline

\multicolumn{3}{l}{Dimension 5 operators}\\[0.5ex] \hline

$\,\checkmark$&$\,g_0\,\bpsi_s \sigma_{\mu \nu} G_{\mu \nu} \psi_d$
& $-$ & + & + & + \\[0.5ex]\hline

$\,$& $\,i\,g_0\,\bpsi_s \gamma_5\sigma_{\mu \nu} G_{\mu \nu} \psi_d$  & +
& + & + & $-$\\[0.5ex]\hline

$\,\checkmark$&$\,(m_{d}^2+m_{s}^2)\bpsi_s\psi_d$ & $-$ & + & +
& + \\[0.5ex]\hline

$\,$& $\,i\,(m_{d}^2+m_{s}^2)\bpsi_s\gamma_5\psi_d$ & + & + & + & $-$
\\[0.5ex]\hline

$\,$& $\,(m_{d}^2-m_{s}^2)\bpsi_s\psi_d$ & $-$ & + & $-$ & $-$
\\[0.5ex]\hline

$\,$& $\,i\,(m_{d}^2-m_{s}^2)\bpsi_s\gamma_5\psi_d$ & + & + & $-$ &
+\\[0.5ex]\hline

$\,\checkmark$&$\,m_{d}\,m_{s}\bpsi_s\psi_d$ & $-$ & + & + & +
\\[0.5ex]\hline

$\,$& $\,i\,m_{d}\,m_{s}\bpsi_s\gamma_5\psi_d$ & + & + & + & $-$ \\[0.5ex]\hline

$\,\checkmark$&$\,m_{s}\bpsi_s(\Dr+m_d)\psi_d
  +
  m_{d}\bpsi_s(-\Dl +m_s)\psi_d$&
$-$ & + & + & + \\[0.5ex]\hline

$\,\checkmark$&$\,m_{d}\bpsi_s(\Dr+m_d)\psi_d
  +
  m_{s}\bpsi_s(-\Dl +m_s)\psi_d$
& $-$ & + & + & +\\[0.5ex]\hline

$\,$& $\,m_{s}\bpsi_s(\Dr+m_d)\psi_d
-
m_{d}\bpsi_s(-\Dl +m_s)\psi_d$
& $-$ & + & $-$ & $-$\\[0.5ex]\hline

$\,$& $\,m_{d}\bpsi_s(\Dr+m_d)\psi_d - 
m_{s}\bpsi_s(-\Dl +m_s)\psi_d$ & $-$ & + & $-$ & $-$\\[0.5ex]\hline

$\,$& $\,i\,m_{s}\bpsi_s\gamma_5(\Dr+m_d)\psi_d + 
i\,m_{d}\bpsi_s(-\Dl +m_s)\gamma_5\psi_d$ & +  & + & + & $-$\\[0.5ex]\hline

$\,$& $\,i\,m_{d}\bpsi_s\gamma_5(\Dr+m_d)\psi_d + 
i\,m_{s}\bpsi_s(-\Dl +m_s)\gamma_5\psi_d$ & +  & + & + & $-$\\[0.5ex]\hline

$\,$& $\,i\,m_{s}\bpsi_s\gamma_5(\Dr+m_d)\psi_d - 
i\,m_{d}\bpsi_s(-\Dl +m_s)\gamma_5\psi_d$& +  & + & $-$ & + \\[0.5ex]\hline

$\,$& $\,i\,m_{d}\bpsi_s\gamma_5(\Dr+m_d)\psi_d - 
i\,m_{s}\bpsi_s(-\Dl +m_s)\gamma_5\psi_d$ & +  & + &  $-$ & +\\[0.5ex]\hline

$\,\checkmark$&$\,\bpsi_s(\Dr+m_d)^2\psi_d + 
\bpsi_s(-\Dl +m_s)^2\psi_d$& $-$ & + & + & + \\[0.5ex]\hline

$\,$& $\,\bpsi_s(\Dr+m_d)^2\psi_d - 
\bpsi_s(-\Dl +m_s)^2\psi_d$ & $-$ & + & $-$ & $-$\\[0.5ex]\hline

$\,$& $\,i\,\bpsi_s\gamma_5(\Dr+m_d)^2\psi_d + 
i\,\bpsi_s(-\Dl +m_s)^2\gamma_5\psi_d$ & + & + & + & $-$\\[0.5ex]\hline

$\,$& $\,i\,\bpsi_s\gamma_5(\Dr+m_d)^2\psi_d - 
i\,\bpsi_s(-\Dl +m_s)^2\gamma_5\psi_d$& + & + & $-$ & + \\[0.5ex]\hline

$\,\checkmark$&$\, \Box \left( \bpsi_s \psi_d \right)$& $-$ & + & + & + \\[0.5ex]\hline

$\,$& $\,i\,\bpsi_s \gamma_5\DDl_\mu \DDr _\mu \psi_d$& + & + & + & $-$ \\[0.5ex]\hline

$\,\checkmark$&$\,\bpsi_s(-\Dl +m_s)\,(\Dr+m_d)\psi_d$& $-$ & + & + & + \\[0.5ex]\hline

$\,$& $\,i\,\bpsi_s(-\Dl +m_s)\,\gamma_5\,(\Dr+m_d)\psi_d$& + & + & + & $-$ \\[0.5ex]\hline

$\,\checkmark$&$\,\bpsi_s  \partl(\Dr +m_d )\psi_d 
  - \bpsi_s (-\Dl +m_s)\partr \psi_d $& $-$ & + & + & + \\[0.5ex]\hline

$\,\checkmark$&$\,\bpsi_s  \partr (\Dr +m_d )\psi_d 
  - \bpsi_s (-\Dl +m_s)\partl\psi_d $& $-$ & + & + & + \\[0.5ex]\hline

$\,$& $\,\bpsi_s \partl(\Dr +m_d )\psi_d 
  + \bpsi_s (-\Dl +m_s)\partr \psi_d $& $-$ & + & $-$ & $-$ \\[0.5ex]\hline

$\,$& $\,\bpsi_s  \partr (\Dr  +m_d )\psi_d 
  + \bpsi_s (-\Dl +m_s)\partl\psi_d $& $-$ & + & $-$ & $-$ \\[0.5ex]\hline

$\,$& $\,i\,\bpsi_s  \partl\gamma_5(\Dr   +m_d )\psi_d 
  - i\,\bpsi_s (-\Dl +m_s)\gamma_5\partr \psi_d $& + & + & + & $-$ \\[0.5ex]\hline

$\,$& $\,i\,\bpsi_s \partr \gamma_5(\Dr +m_d )\psi_d 
  - i\,\bpsi_s (-\Dl +m_s)\gamma_5\partl\psi_d $& + & + & + & $-$ \\[0.5ex]\hline

$\,$& $\,i\,\bpsi_s \partl\gamma_5(\Dr +m_d )\psi_d 
  + i\,\bpsi_s (-\Dl +m_s)\gamma_5\partr \psi_d $ & + & + & $-$ & +\\[0.5ex]\hline

$\,$& $\,i\,\bpsi_s \partr \gamma_5(\Dr +m_d )\psi_d
  + i\,\bpsi_s (-\Dl +m_s)\gamma_5\partl\psi_d $& + & + & $-$ & + \\[0.5ex]\hline

\end{tabular}
\caption{Transformation properties of gauge invariant operators and of operators which vanish by the equations of motion, in the physical basis.}
\label{tb:GaugeVariantAnde.m.o}
\end{center}
\end{table}

In Table~\ref{tb:GaugeVariantAnde.m.o} we present all candidate
operators along with their transformation properties. Operators marked
by $''\checkmark''$ have the same properties as ${\cal O}_{CM}$ and thus
may mix with it. Operators marked by $''(+)''$ or $''(-)''$ have the same
transformation properties as ${\cal O}_{CM}$ only if $r_s=r_d$ or
$r_s=-r_d$, respectively; for this reason the Wilson parameters
$r_s,\,r_d$ have been explicitly introduced in ${\cal O}_{11}$ and
${\cal O}_{12}$ below (see Eqs.~(\ref{O11}) - (\ref{O12})). There follows immediately that ${\cal
 O}_{CM}\equiv {\cal O}_1$ can only mix with the following operators:
\bea
\label{O1}
{\cal O}_1    &=& g_0\,\bpsi_s \sigma_{\mu \nu} G_{\mu \nu} \psi_d  \\[1ex]
{\cal O}_2    &=&  (m_{d}^2+m_{s}^2) \bpsi_s\psi_d \\[1ex]
{\cal O}_3    &=&  m_{d}\,m_{s} \bpsi_s\psi_d \\[1ex]
{\cal O}_4    &=&  \Box  \left( \bpsi_s \psi_d \right) \\[1ex]
{\cal O}_5    &=& \bpsi_s (-\Dl +m_s)(\Dr+m_d)\psi_d \\[1ex]
{\cal O}_6    &=& \bpsi_s (\Dr+m_d)^2\psi_d + \bpsi_s(-\Dl +m_s)^2\psi_d \\[1ex]
{\cal O}_7    &=& m_{s} \bpsi_s(\Dr+m_d)\psi_d + m_{d}\bpsi_s(-\Dl +m_s)\psi_d \\[1ex]
{\cal O}_8    &=& m_{d} \bpsi_s(\Dr+m_d)\psi_d + m_{s}\bpsi_s(-\Dl +m_s)\psi_d \\[1ex]
{\cal O}_9    &=&  \bpsi_s \partl(\Dr+m_d)\psi_d - \bpsi_s (-\Dl +m_s)\partr \psi_d \\[1ex]
{\cal O}_{10} &=& \bpsi_s \partr (\Dr+m_d)\psi_d - \bpsi_s (-\Dl +m_s)\partl\psi_d \\[1ex]
\label{O11} 
{\cal O}_{11} &=& i\,r_d\,\bpsi_s\gamma_5(\Dr+m_d)\psi_d + i\,r_s\,\bpsi_s(-\Dl +m_s)\gamma_5\psi_d \\[1ex]
\label{O12}
{\cal O}_{12} &=& i\,(r_d\,m_{d}+r_s\,m_{s})\bpsi_s\gamma_5\psi_d  \\[1ex]
{\cal O}_{13} &=& \bpsi_s\,\psi_d\,,
\label{O13}
\eea
where $\Box \equiv \partial_\mu \partial_\mu$, and the left and right covariant derivatives are defined in terms of the gluon field $A_\mu$ as follows:
\be
\DDr_\mu = \overrightarrow{\partial}_\mu + i g_0 A_\mu ~, \qquad \qquad \DDl_\mu = \overleftarrow{\partial}_\mu - i g_0 A_\mu ~ .
\ee

We do not impose in our calculation the conservation of external momentum. Therefore, the list of independent operators in Table~\ref{tb:GaugeVariantAnde.m.o} accounts for operators which are total derivatives\footnote{Instead of the operator ${\cal O}_4$ one can consider the operator ${\cal O}_4^\prime \equiv \bpsi_s \DDl_\mu \DDr _\mu \psi_d$, which shares the same transformation properties of ${\cal O}_4$. It can easily be shown that the operator  ${\cal O}_4^\prime$ is a linear combination of ${\cal O}_4$ and other operators entering the basis (\ref{O1}-\ref{O13}). The choice of ${\cal O}_4$ has the advantage that it does not contribute to physical amplitudes, since it is a total four-derivative. Moreover, its non-perturbative determination is quite simpler, because its matrix element between physical hadron states is simply given by the corresponding matrix element of the scalar density $\bpsi_s \psi_d$ multiplied by the squared four-momentum transfer. Finally, as it will be shown later, its mixing with the chromomagnetic operator is vanishing at one-loop and therefore any redefinition of ${\cal O}_4$ does not change the mixing of the other operators of the basis at one-loop.}.
As for the parameters $r_s$ and $r_d$, in our perturbative calculation we make the (independent) choices of values $r_s = \pm 1$,\, $r_d = \pm 1$, consistently with their values in the simulations.

Operators ${\cal O}_{9}$ and ${\cal O}_{10}$ are not gauge invariant, but they are admissible candidates for mixing, since they vanish by the equations of motion; indeed, they will mix with ${\cal O}_{CM}$ both in DR and on the lattice.
The operators ${\cal O}_{11},\,{\cal O}_{12},\,{\cal O}_{13}$ are of lower dimension and thus they do not mix with ${\cal O}_{1}$ in DR; they do however show up in the lattice formulation. 

Before closing this Section we mention that in the presence of the electromagnetic interactions the operator ${\cal{O}}_{CM}$ can mix also with the EMO (see Eq. (\ref{Qgammapm})).
The corresponding mixing coefficient is of order ${\cal{O}}(g^2)$, i.e.~it does not vanish formally in the limit of zero quark electric charge, since the latter is already included in the definition of the EMO.

\section{Renormalization functions}
\label{RFs}

The operators ${\cal O}^R_i$ are related to the bare ones, ${\cal O}_i$ ($i=1,\ldots,13$), through:
\begin{equation}
{\cal O}_i = \sum_{j=1}^{13} Z_{ij} {\cal O}_j^R\,\,\quad ({\rm in\,\, matrix\,\, notation:}\,\, {\cal O}=Z {\cal O}^R )\,,
\end{equation}
where the $13 \times 13$ mixing matrix $Z_{ij}$ (which should more properly be denoted as $Z_{ij}^{X,Y}$, where $X = DR, L, \ldots$ is the regularization and $Y = \overline{MS}, RI^\prime, \ldots$ the renormalization scheme) obeys:
\begin{equation}
Z = \openone + {\cal O}(g^2)\,,
\label{Z.eq28}
\end{equation}
where $g$ is the renormalized coupling constant.
Since we are interested in ${\cal O}_1^R$ we only need to calculate the first row of the inverse mixing matrix, $Z^{-1}$, which, to one loop, is immediately related to the first row of $Z$: $Z_i \equiv Z_{1i}$. 

\begin{figure}[htb!]
\centering
\includegraphics[scale=0.5]{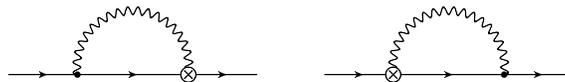}
\caption{One-loop Feynman diagrams contributing to the 2-pt Green's
  function of the chromomagnetic operator, ${\cal O}_1$\,.  A wavy (solid) line represents gluons
  (quarks). 
A crossed circle denotes the insertion of ${\cal O}_1$\,.}
\label{fig2pt}
\end{figure}

Since renormalization conditions are typically imposed on amputated renormalized Green's functions, let us relate the latter to the bare ones.
For the quark-quark Green's functions:
\bea
\langle \psi^R\,{\cal O}_1^R\,\bpsi^R \rangle_{\rm amp} &=& 
\langle\psi^R\,\bpsi^R \rangle^{-1} \,
\langle\psi^R\,{\cal O}_1^R\,\bpsi^R \rangle \,
\langle\psi^R\,\bpsi^R \rangle^{-1} \nonumber \\
&=& \left( Z_\psi\,\langle\psi\,\bpsi \rangle^{-1}\right) \,
\Big(Z^{-1}_\psi\,\sum_{i=1}^{13}(Z^{-1})_{1i}\langle\psi\,{\cal O}_i\,\bpsi \rangle \Big)\,
\left(Z_\psi\,\langle\psi\,\bpsi \rangle^{-1}\right) \nonumber \\
&=& Z_\psi\,\sum_{i=1}^{13}(Z^{-1})_{1i} \langle\psi\,{\cal O}_i\,\bpsi \rangle _{\rm amp}\,,\quad \psi=\sqrt{Z_\psi}\,\psi^R\,.
\label{GG2}
\eea
The one-loop Feynman diagrams contributing to $\langle\psi\,{\cal O}_1\,\bpsi
\rangle _{\rm amp}$ are shown in Fig.~\ref{fig2pt}. Note that Eq.~(\ref{GG2}) holds for an arbitrary regularization and
arbitrary renormalization scheme; the only condition on the renormalization scheme is that it be mass-independent, in
which case the quark field renormalization constant $Z_\psi$ does not depend on flavor. To avoid heavy notation we 
have omitted coordinate/momentum arguments on $\psi,\,{\cal O}$, as well as Dirac/flavor indices on $\langle\psi\,\bpsi \rangle, \langle\psi\,{\cal O}\,\bpsi\rangle$, etc.

Similarly for quark-quark-gluon Green's functions we have:
\be
\langle \psi^R\,{\cal O}_1^R\,\bpsi^R A_\nu^R \rangle_{\rm amp} =
Z_\psi\,Z_A^{1/2}\sum_{i=1}^{13}(Z^{-1})_{1i} \langle\psi\,{\cal
  O}_i\,\bpsi\,A_\nu \rangle _{\rm amp}\,,\quad  A_{\nu}= \sqrt{Z_A}\,A^{R}_{\nu}\,.
\label{GG3}
\ee
Strictly speaking, on the right-hand sides of Eqs.~(\ref{GG2}) and~(\ref{GG3}) one must take the regulator to its limit
value (i.e. $\epsilon \to 0$ in DR or $a \to 0$ on the lattice). This limit is convergent, provided
all renormalization functions $Z$ have been appropriately chosen. It is only in this limit that the right-hand sides of Eqs.~(\ref{GG2}) and~(\ref{GG3}) are equal to the corresponding left-hand sides.

\begin{figure}[htb!]
\centering
\includegraphics[scale=0.7]{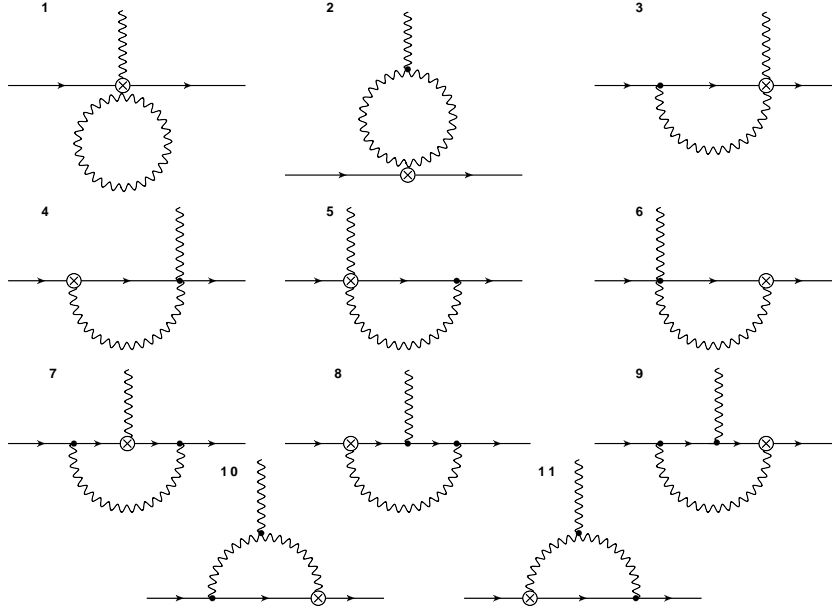}
\caption{1PI Feynman diagrams which contribute to the 3-pt Green's function of ${\cal O}_1$. Diagrams 1, 4, 6 do not appear in DR.
Wavy (solid) lines represent gluons (quarks). Crossed circles denote the insertions of ${\cal O}_1$\,.}
\label{fig3pt1PI}
\end{figure}

\begin{figure}[htb!]
\centering
\includegraphics[scale=0.5]{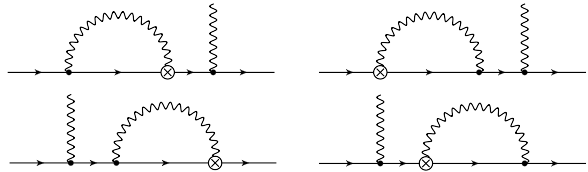}
\caption{1PR Feynman diagrams which contribute to the 3-pt Green's function of ${\cal O}_1$.
Wavy (solid) lines represent gluons (quarks). Crossed circles denote the insertions of ${\cal O}_1$\,.}
\label{fig3pt1PR}
\end{figure}

The one-loop Feynman diagrams contributing to $\langle\psi\,{\cal O}_1\,\bpsi
\,A_\nu \rangle _{\rm amp}$ are shown in Fig.~\ref{fig3pt1PI}
(one-particle irreducible (1PI)) and Fig.~\ref{fig3pt1PR} (one-particle reducible (1PR)). While 1PR diagrams
do contribute to the renormalized Green's functions $\langle \psi^R\,{\cal O}_1^R\,\bpsi^R A_\nu^R \rangle_{\rm amp}$,
their contribution to the mixing matrix cancels out.

Imposing renormalization conditions of the above 2- and 3-pt Green's functions is sufficient\footnote{One could of course calculate also 4-pt Green's functions; in doing so, a number of consistency checks would emerge regarding the divergent part of the mixing coefficients $Z_i$. Further Green's functions (5-pt and above) will bring in no superficial divergences, and no further renormalization conditions (or consistency checks) will arise.} in order to obtain all $Z_i$.

In some definitions of ${\cal O}_{CM}$ (see, e.g., \cite{Ciuchini:1993fk})
there is an extra factor of a quark mass:
\be
\widetilde{\cal O}_{CM} \equiv m\,{\cal O}_{CM}\,,
\label{Otilde2}
\ee
where $m$ is the mass of one of the quark flavors. 
The renormalized mass $m^R$ is given by $m^R=Z^{-1}_m\,m$; in a mass-independent scheme, $Z_m$ is also flavor independent, like $Z_\psi$. 
In this case the renormalization matrix $\tilde Z_{ij}$ for $\tilde{\cal O}_{CM}$ is simply given by: $\tilde Z_{ij}=Z_m\,Z_{ij}$.

By analogy with $Z_m$, a multiplicative factor of $Z_g$ must be included in $Z_{1}$, if the calculation of Green's functions involves the operator $\bpsi_s\,\sigma_{\mu\nu}\,G_{\mu\nu}\,\psi_d$, rather than $g\,\bpsi_s\,\sigma_{\mu\nu}\,G_{\mu\nu}\,\psi_d$. 
We will make use of this fact in Eq.~(\ref{eqz1}).
The calculation of $Z_m$ and $Z_g$ is presented in Appendix~\ref{app:Z}.

\begin{figure}[htb!]
\centering
\includegraphics[scale=0.6]{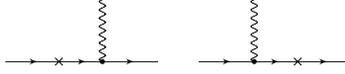}
\caption{1PR Feynman diagrams contributing to the tree-level 3-pt Green's functions. Wavy (solid) lines represent gluons (quarks). Crosses denote the insertions of the operator ${\cal O}_i\,,\, i = 1, \ldots ,13$.}
\label{fig3ptTree}
\end{figure}

In order to impose renormalization conditions, we need the expressions for the tree-level 2-pt and 3-pt Green's functions of ${\cal O}_i$, $i=1,\ldots,13$.  
The tree-level parts of the 3-pt amputated bare Green's functions $\langle \psi_s(q_2) {\cal O}_{i}(x) \bpsi_d(q_3)\,A_\nu(q_1)\rangle_{\rm amp}$ are shown (apart
from an overall factor of $e^{i\,x\cdot(-q_1-q_2+q_3)}$) in Table \ref{tb:mixing_operators}; similarly for the tree-level parts of the 2-pt bare Green's functions  $\langle \psi_s(q_2) {\cal O}_{i}(x) \bpsi_d(q_3)\rangle_{\rm amp}$. 
Note that the tree-level 3-pt Green's functions, despite being amputated, receive also contributions which are not 1PI, as shown in Fig.~\ref{fig3ptTree}. 
We do not include these in Table \ref{tb:mixing_operators}; however, their value can be easily deduced from the corresponding tree-level 2-pt Green's functions.

\begin{sidewaystable}
\begin{center}
\begin{tabular}{lr@{}lr@{}lr@{}l}
\hline
\multicolumn{1}{c}{$\,\,\,$$\,\,\,$$\,\,\,$$\,\,\,$$\,\,\,$}&
\multicolumn{2}{c}{$\,\,\,$$\,\,\,$Operators                  $\,\,\,$$\,\,\,$}&
\multicolumn{2}{c}{$\,\,\,$$\,\,\,$Tree Level 2-pt$\,\,\,$$\,\,\,$}&
\multicolumn{2}{c}{$\,\,\,$$\,\,\,$$\,\,\,$$\,\,\,$$\,\,\,$$\,\,\,$$\,\,\,$$\,\,\,$$\,\,\,$$\,\,\,$$\,\,\,$$\,\,\,$$\,\,\,$$\,\,\,$$\,\,\,$$\,\,\,$$\,\,\,$$\,\,\,$Tree
  Level 3-pt (1PI)$\,\,\,$$\,\,\,$$\,\,\,$$\,\,\,$$\,\,\,$$\,\,\,$$\,\,\,$$\,\,\,$$\,\,\,$$\,\,\,$$\,\,\,$$\,\,\,$$\,\,\,$$\,\,\,$$\,\,\,$}\\
\hline
\hline
$\,{\cal O}_1\, $ $\,\,\,$&$g_0\bpsi_s \sigma_{\mu \nu} G_{\mu \nu} \psi_d$&$\,\,\,$       $\,\,\,$&0&$\,\,\,$&$-2 i g_0 \sigma_{\mu \nu} q_{1\mu}$ \\[1.2ex]\hline

$\,{\cal O}_2\, $ $\,\,\,$&$(m_{d}^2+m_{s}^2)\bpsi_s\psi_d$&$\,\,\,$       $\,\,\,$&$m_{d}^2+m_{s}^2$&$\,\,\,$&0 \\[1.2ex]\hline

$\,{\cal O}_3\, $ $\,\,\,$&$m_{d}m_{s}\bpsi_s\psi_d$&$\,\,\,$       $\,\,\,$&$m_{d}m_{s}$&$\,\,\,$&0 \\[1.2ex]\hline

$\,{\cal O}_4\, $ $\,\,\,$&$\Box \left( \bpsi_s \psi_d \right)$&$\,\,\,$       $\,\,\,$&$(q_2 - q_3)^2$&$\,\,\,$&$0$ \\[1.2ex]\hline

$\,{\cal O}_5\, $ $\,\,\,$&$\bpsi_s(-\Dl +m_s)(\Dr+m_d)\psi_d$&$\,\,\,$       $\,\,\,$&$-\qq \qqq+i\qq m_d + i\qqq m_s + m_sm_d$&$\,\,\,$& $-g_0(\qq\gamma_\nu+\gamma_\nu\qqq )+ ig_0(m_s+m_d)\gamma_\nu$\\[1.2ex]\hline

$\,{\cal O}_6\, $ $\,\,\,$&$\bpsi_s(\Dr+m_d)^2\psi_d + \bpsi_s(-\Dl +m_s)^2\psi_d$&$\,\,\,$       $\,\,\,$&$-q_2^2-q_3^2+2i(m_d\qqq+m_s\qq)+m_{d}^2+m_{s}^2$&$\,\,\,$&$-2g_0 i \sigma_{\mu \nu} q_{1\mu}- 2g_0(q_{3\nu}+q_{2\nu}) -2ig_0(m_d+m_s)\gamma_\nu$ \\[1.2ex]\hline

$\,{\cal O}_7\, $ $\,\,\,$&$m_{s}\bpsi_s(\Dr+m_d)\psi_d + m_{d}\bpsi_s(-\Dl +m_s)\psi_d$&$\,\,\,$       $\,\,\,$&$m_s(\
i\qqq+m_d)+m_d(i\qq+m_s)$&$\,\,\,$& $ig_0 (m_s+m_d)\gamma_\nu$\\[1.2ex]\hline

$\,{\cal O}_8\, $ $\,\,\,$&$m_{d}\bpsi_s(\Dr+m_d)\psi_d + m_{s}\bpsi_s(-\Dl +m_s)\psi_d$&$\,\,\,$       $\,\,\,$&$m_d(\
i\qqq+m_d)+m_s(i\qq+m_s)$&$\,\,\,$&$ig_0 (m_s+m_d)\gamma_\nu$ \\[1.2ex]\hline

$\,{\cal O}_9\, $ $\,\,\,$&$\bpsi_s                                                                                                                                                                    
  \partl(\Dr                                                                                                                   
  +m_d )\psi_d                                                                                                                                                                                            
  - \bpsi_s (-\Dl +m_s)\partr \psi_d $&$\,\,\,$       $\,\,\,$&$2 \qq \qqq        
- i (\qq m_d + \qqq m_s)$&$\,\,\,$&$g_0(\qq\gamma_\nu+\gamma_\nu\qqq)$ \\[1.2ex]\hline

$\,{\cal O}_{10}\, $ $\,\,\,$&$\bpsi_s                                                                                                                                                                 
  \partr (\Dr                                                                                                                  
  +m_d )\psi_d                                                                                                                                       
  - \bpsi_s (-\Dl +m_s)\partl\psi_d $&$\,\,\,$       $\,\,\,$&$-q_3^2-q_2^2+i (\qq m_s + \qqq m_d)$&$\,\,\,$&$-2g_0(q_{2\nu}+q_{3\nu})+g_0(\qq\gamma_\nu+\gamma_\nu\qqq) - 2 i g_0 \sigma_{\mu \nu} q_{1\mu}$ \\[1.2ex]\hline

$\,{\cal O}_{11}\, $ $\,\,\,$&$i\,r_d\bpsi_s\gamma_5(\Dr+m_d)\psi_d +i\,r_s\bpsi_s(-\Dl +m_s)\gamma_5\psi_d$&$\,\,\,$       \
$\,\,\,$&$i\,r_d\gamma_5(i\qqq+m_d)+i\,r_s(i\qq+m_s)\gamma_5$&$\,\,\,$&$-g_0 (r_d-r_s)\gamma_5\gamma_\nu $\\[1.2ex]\hline

$\,{\cal O}_{12}\, $ $\,\,\,$&
$i\,(r_d\,m_{d}+r_s\,m_{s})\bpsi_s\gamma_5\psi_d$&$\,\,\,$       $\,\,\,$&$i\,(r_d\,m_{d}+r_s\,m_{s})\gamma_5$&$\,\,\,$&0 \\[1.2ex]\hline

$\,{\cal O}_{13}\, $ $\,\,\,$&$\bpsi_s\psi_d$&$\,\,\,$       $\,\,\,$&1&$\,\,\,$&0 \\\hline

\hline
\end{tabular}
\end{center}
\caption{Operators which will possibly mix with the chromomagnetic operator in the physical basis, along with their tree-level 2-pt and 3-pt (1PI) Green's functions. Here, $q_1$ is the external gluon momentum and $q_{2, 3}$ is the external final (initial) quark momentum.}
\label{tb:mixing_operators}
\end{sidewaystable}

\subsection{Dimensional Regularization}
\label{sec2.1}

The next step in our renormalization procedure is to calculate the $\MSbar$-renormalized 2-pt and 3-pt Green's functions of  ${\cal O}_{CM}$\,; in order to do so, we must regularize the theory in D-dimensions ($D = 4 - 2\,\epsilon$), in the continuum.
The general form of the ${\cal O}\left(1/\epsilon \right)$ part of the bare Green's functions is (consistently with the tree-level values of the operators in Table \ref{tb:mixing_operators}):
\bea
\label{G2}
\langle \psi\,{\cal O}_1\,\bpsi \rangle^{DR}_{\rm amp}\Big{|}_{1/\epsilon} \hspace{-0.1cm}&=&\hspace{-0.1cm} \rho_1\,(q^2_2 + q^2_3) + \rho_2\,(m_s^2 + m^2_d) + \rho_3\,i\,(m_d\,\qqq + m_s\,\qq) \nonumber \\[1ex]
&&+ \rho_4\,i\,(m_s\,\qqq + m_d\,\qq) + \rho_5\,q_2.q_3 + \rho_6\,\qq\,\qqq + \rho_7\,m_s\,m_d\qquad
\\[3ex]
\label{G3}
\langle \psi\,{\cal O}_1\,\bpsi A_\nu \rangle^{DR}_{{\rm amp}, 1PI}\Big{|}_{1/\epsilon} &=& R_1\,g \,(q_2+q_3)_\nu + R_2\,g\,(\gamma_\nu\,\qqq+ \qq\,\gamma_\nu) + 
R_3\,i\,g\,(m_s+m_d)\,\gamma_\nu + R_4\,(-2\,i\,g\,\sigma_{\rho\nu}\,q_{1\rho}\,)
\eea
where $g$ is the renormalized coupling constant in the $\overline{\rm MS}$ scheme, which is related to the bare coupling constant in DR, $g_0^{DR}$ , through: $g=\mu^{-\epsilon}\,(Z^{DR,\MSbar}_g)^{-1}\,g_0^{DR}$ and $\rho_i,\,R_i$ are numerical coefficients. 
Additional terms in Eqs.~(\ref{G2}) and~(\ref{G3}) (such as $gq_{1 \nu}$ or $g \qslash_1 \gamma_{\nu}$ in Eq.~(\ref{G3})) would imply mixing with further operators from Table \ref{tb:GaugeVariantAnde.m.o}, but this is excluded by the symmetries listed in Eq.~(\ref{symmetries}).

Computing $\rho_i,\,R_i$ to one loop we find:
\bea
\label{r1}
\rho_1 &=& \frac{g^2}{16\,\pi^2} \frac{1}{\epsilon} \left( -3 C_F \right)\\[1ex]
\rho_2 &=& \frac{g^2}{16\,\pi^2} \frac{1}{\epsilon} \left( -6 C_F \right)\\[1ex]
\rho_3 &=& \frac{g^2}{16\,\pi^2} \frac{1}{\epsilon} \left( 3 C_F \right)\\[1ex]
\rho_4 &=& \rho_5=\rho_6=\rho_7=0 \phantom{ \frac{g^2\,C_F}{16\,\pi^2}\left(-\frac{3}{\epsilon} \right)}
\eea
\bea
\label{RR1}
R_1 &=& \frac{g^2}{16\,\pi^2} \frac{1}{\epsilon} \left( -6 C_F \right)\\[1ex]
R_2 &=& \frac{g^2}{16\,\pi^2} \frac{1}{\epsilon} \left(\frac{3\,N_c}{4} \right) \\[1ex]
R_3 &=& \frac{g^2}{16\,\pi^2} \frac{1}{\epsilon} \left(-\frac{3}{2\,N_c} +\frac{3\,N_c}{4} \right) \\[1ex]
R_4 &=& \frac{g^2}{16\,\pi^2} \frac{1}{\epsilon} \left(\frac{1}{N_c}-\frac{\alpha}{2\,N_c}+\frac{7\,N_c}{4}+\frac{3\,\alpha\,N_c}{4} \right)\,.
\eea
Here, $N_c$ is the number of colors, $C_F=(N_c^2-1)/(2\,N_c)$ is the quadratic Casimir operator in the fundamental representation, $\alpha$ is the gauge parameter ($\alpha=1(0)$ corresponds to Feynman (Landau) gauge).

We have also computed the finite parts (${\cal O}(\epsilon^0)$) for the above Green's
functions, which are just the corresponding $\MSbar$-renormalized Green's
functions. These are irrelevant for the computation of the mixing
coefficients in the $\MSbar$ scheme in DR; however, they are necessary in the calculation of
$Z_{ij}$ with lattice regularization and $\MSbar$
renormalization, see Section~\ref{sec2}. Using the form of
Eqs.~(\ref{G2}) - (\ref{G3}) and the tree-level Green's functions of
the various operators (Table \ref{tb:mixing_operators}), we construct a set of equations for
the disentanglement of the mixing coefficients; in particular, by
demanding that the coefficients of ${\cal O}(1/\epsilon)$ in the
left-hand sides of Eqs.~(\ref{GG2}) - (\ref{GG3}) vanish\footnote{Note that Eq.~(\ref{GG3}) 
will also contain ${\cal O}\left(1/\epsilon \right)$ terms which are not polynomial in $q_i$, $m$; 
such terms arise from the 1PR one-loop 3-pt Green's function of ${\cal O}_1$ (Fig.~\ref{fig3pt1PR}) and from the 
1PR tree-level Green's functions of ${\cal O}_2, \ldots, {\cal O}_{13}$ (Fig.~\ref{fig3ptTree}). By Eq.~(\ref{GG2}) 
all such terms cancel out among themselves.}, we obtain, to order $g^2$:
\bea
\label{eq1}
-\,Z^{DR,\MSbar}_6-Z^{DR,\MSbar}_{10} &=& \rho_1  \\[1ex]
Z^{DR,\MSbar}_2+Z^{DR,\MSbar}_6+Z^{DR,\MSbar}_8 &=& \rho_2 \\[1ex]
2\,Z^{DR,\MSbar}_6+Z^{DR,\MSbar}_8+Z^{DR,\MSbar}_{10} &=& \rho_3 \\[1ex]
-\,Z^{DR,\MSbar}_5-Z^{DR,\MSbar}_7+Z^{DR,\MSbar}_9 &=& \rho_4 \\[1ex]
\label{eq5}
-\,Z^{DR,\MSbar}_4 &=& \rho_5 \\[1ex]
-\,Z^{DR,\MSbar}_5+2\,Z^{DR,\MSbar}_9 &=& \rho_6  \\[1ex]
-\,Z^{DR,\MSbar}_3-Z^{DR,\MSbar}_5-2\,Z^{DR,\MSbar}_7 &=& \rho_7  \\[1ex]
\label{eq9}
Z^{DR,\MSbar}_4-2\,Z^{DR,\MSbar}_6-2\,Z^{DR,\MSbar}_{10} &=& R_1 \\[1ex]
-\,Z^{DR,\MSbar}_5+Z^{DR,\MSbar}_9+Z^{DR,\MSbar}_{10} &=& R_2 \\[1ex]
\label{eq11}
Z^{DR,\MSbar}_5-2\,Z^{DR,\MSbar}_6+Z^{DR,\MSbar}_7+Z^{DR,\MSbar}_8 &=& R_3 \\[1ex]
\label{eqz1}
g^2\,z^{DR,\MSbar}_1 + Z^{DR,\MSbar}_6+Z^{DR,\MSbar}_{10} &=& R_4 + g^2 \left( z^{DR,\MSbar}_\psi + \frac{1}{2}\,z^{DR,\MSbar}_A + z^{DR,\MSbar}_g \right)
\eea
where  $Z^{DR,\MSbar}_1= 1+ g^2\,z^{DR,\MSbar}_1 + {\cal O}(g^4)$ and $Z_{i>1} = {\cal O}(g^2)$ by Eq.~(\ref{Z.eq28}); also:
\bea
Z^{DR,\MSbar}_\psi=1+g^2\,z^{DR,\MSbar}_\psi+{\cal O}(g^4)\,,\qquad z^{DR,\MSbar}_\psi &=&   \frac{1}{16\,\pi^2}\frac{1}{\epsilon}\left(-C_F\,\alpha \right) \\[1ex]
Z^{DR,\MSbar}_A=1+g^2\,z^{DR,\MSbar}_A+{\cal O}(g^4)\,,      \qquad z^{DR,\MSbar}_A    &=&   \frac{1}{16\,\pi^2}\frac{1}{\epsilon}\left(\frac{13\,N_c}{6} -\frac{\alpha\,N_c}{2} -\frac{2\,N_f}{3}\right)\\[1ex]
Z^{DR,\MSbar}_g=1+g^2\,z^{DR,\MSbar}_g + {\cal O}(g^4)\,,    \qquad z^{DR,\MSbar}_g    &=&   \frac{1}{16\,\pi^2}\frac{1}{\epsilon}\left(\frac{N_f}{3} -\frac{11\,N_c}{6}\right)\,.
\eea
In particular, Eq.~(\ref{eqz1}) stems from the requirement that the coefficients of $(-2 i g \sigma_{\mu \nu} q_{1\mu}) / \epsilon$ in the left-hand side and right-hand side of Eq.~(\ref{GG3}) coincide:
\be
0=\underbrace{(1+g^2\,z^{DR,\MSbar}_\psi)\,\left(1+\frac{1}{2}\,g^2\,z^{DR,\MSbar}_A\right)\,\left(1+g^2\,z^{DR,\MSbar}_g\right)\,(1-g^2\,z^{DR,\MSbar}_1)(1+R_4)}_{{\rm only\,\,the}\,\, {\cal O}(1/\epsilon)\,\,{\rm part}}\, - \,Z^{DR,\MSbar}_6 - Z^{DR,\MSbar}_{10}\,.
\ee
As it stands, the system of 11 equations (Eq.~(\ref{eq1}) - (\ref{eqz1})) 
for the 10 unknowns $Z^{DR,\MSbar}_1 - Z^{DR,\MSbar}_{10}$ appears over-constrained; indeed,
Eqs. (\ref{eq1}), (\ref{eq5}) and (\ref{eq9}) can only be compatible
if $2\,\rho_1 = R_1$. This relation is indeed confirmed by our
results (Eq. (\ref{r1}) and Eq. (\ref{RR1})). The presence of $z^{DR,\MSbar}_g$ in
Eq. (\ref{eqz1}) stems from the fact that all one-loop Green's
functions were calculated with an insertion of 
$\bpsi_s\,\sigma_{\mu\nu}\,G_{\mu\nu}\,\psi_d$ 
(rather than $g\,\bpsi_s\,\sigma_{\mu\nu}\,G_{\mu\nu}\,\psi_d$\,; thus multiplication by $Z_g$ is necessary in a way analogous to Eq.~(\ref{Otilde2}).

Solving the above equations, we obtain the mixing coefficients:
\bea
Z^{DR,\MSbar}_1  &=& 1+ \gtilde\,\frac{1}{\epsilon}\,\left(-\frac{N_c}{2}+\frac{5}{2\,N_c}\right) \\[1ex]
Z^{DR,\MSbar}_2  &=& \gtilde\,\frac{1}{\epsilon}\,\left(-3\,N_c+\frac{3}{N_c}\right) \\[1ex]
Z^{DR,\MSbar}_3  &=& 0 \\[1ex]
Z^{DR,\MSbar}_4  &=& 0 \\[1ex]
Z^{DR,\MSbar}_5  &=& \gtilde\,\frac{1}{\epsilon}\,\left(\frac{3\,N_c}{2}-\frac{3}{N_c}\right) \\[1ex]
Z^{DR,\MSbar}_6  &=& 0 \\[1ex]
Z^{DR,\MSbar}_7  &=& \gtilde\,\frac{1}{\epsilon}\,\left(-\frac{3\,N_c}{4} + \frac{3}{2\,N_c}\right) \\[1ex]
Z^{DR,\MSbar}_8  &=& 0 \\[1ex]
Z^{DR,\MSbar}_9  &=& \gtilde\,\frac{1}{\epsilon}\,\left(\frac{3\,N_c}{4}-\frac{3}{2\,N_c}\right) \\[1ex]
Z^{DR,\MSbar}_{10} &=& \gtilde\,\frac{1}{\epsilon}\,\left(\frac{3\,N_c}{2}-\frac{3}{2\,N_c}\right).
\eea

An immediate check of our results is the  extraction of the correct
anomalous dimension, $\widetilde\gamma_{CM}$, already known in the
literature for the operator $\widetilde{\cal O}_{CM}$
(Eq.~(\ref{Otilde2})), with a quark mass and a coupling constant in
its definition~\cite{Ciuchini:1993fk}. The following relation holds between
$z^{DR,\MSbar}_1$ and $\widetilde\gamma_{CM}$:
\be
\widetilde\gamma_{CM} = -2\,\epsilon\,g^2\,(z^{DR,\MSbar}_1 + z^{DR,\MSbar}_m) = \frac{g^2}{16\,\pi^2}\left(4\,N_c -\frac{8}{N_c} \right),
\ee
\be
\left(Z^{DR,\MSbar}_m = 1+ g^2\,z^{DR,\MSbar}_m + {\cal O}(g^4),\quad
z^{DR,\MSbar}_m = \frac{1}{16\,\pi^2}\frac{1}{\epsilon}\left(-3\,C_F \right)\, \right).
\ee

\subsection{Lattice regularization - $\MSbar$ renormalization}
\label{sec2}

The computation of the 2-pt and 3-pt bare Green's functions of ${\cal O}_{CM}$ on the lattice are the most
demanding part of the present work. This is particularly true for the 3-pt function, since it had to be 
calculated for arbitrary values of the external momenta, $q_i$\,, of the quarks and gluon. The 
algebraic expressions involved were split into two parts: a) Terms that can be evaluated in the $a \to 0$
limit: Such terms exhibit a very complicated dependence on $q_i$\,, even for zero quark masses, involving 
Spence functions. These functions constitute a part of the regularization independent renormalized Green's 
functions. b) All remaining terms: These are divergent as $a \to 0$, however their dependence on $q_i, m$ is necessarily
polynomial. Our computations were performed in a covariant gauge, with arbitrary value of the gauge parameter $\alpha$. Given that some of the operators which mix with ${\cal O}_{CM}$ contain powers of the quark masses, we have kept these masses different from zero throughout most of the computation; it is only in the final expressions for $Z_i$ that we set $m \to 0$. 

For the algebraic operations involved in evaluating Feynman diagrams, we make use of
our symbolic package in Mathematica. A brief description of the computation of a Feynman
diagram can be found, e.g., in Ref.~\cite{Constantinou:2009tr} and references therein.
The algebraic expressions for each Feynman diagram typically involve $\sim 10^5$ terms at intermediate stages. The requirements in terms of CPU time, both for algebraic manipulation and for numerical integration of momentum loop integrals, were rather modest as compared to human effort: a total of $\sim 4$ months on a single core CPU was required.

The computation on the lattice is performed in the twisted basis ($\psi',\,\bar{\psi}'$), and
thus, before comparing with the results in DR,
we must rotate to the physical basis ($\psi,\,\bar\psi$). This rotation amounts to the following
transformation of the fermion field:
\bea
\psi' &=& e^{-i\,\frac{\pi}{4}\,\gamma_5}\,\psi\,, \\
\bar \psi' &=& e^{-i\,\frac{\pi}{4}\,\gamma_5}\,\bar \psi\,.
\eea 
The rotation of the 2-pt Green's function is therefore:
\be
\langle \psi\,{\cal O}\,\bar \psi\rangle_{\rm amp} = 
e^{-i\,\frac{\pi}{4}\,\gamma_5}\,\langle\psi'\,{\cal O}\,\bar \psi'\rangle_{\rm amp}\,e^{-i\,\frac{\pi}{4}\,\gamma_5}\,,
\ee
and similarly for the 3-pt Green's function.

We will make use, once again, of Eqs.~(\ref{GG2}) - (\ref{GG3}), with $\MSbar$ being the renormalization
scheme; however, the regularization will now be the lattice. The above equations now take the form:
\be
\langle \psi\,{\cal O}_1\,\bpsi \rangle_{\rm amp}^{\MSbar} = 
Z_\psi^{L,\MSbar}\,\sum_{i=1}^{13}\big((Z^{L,\MSbar})^{-1}\big)_{1i} \langle\psi\,{\cal O}_i\,\bpsi \rangle^{L}_{\rm amp}
\label{GG2lattice}
\ee
and 
\be
\langle \psi\,{\cal O}_1\,\bpsi A_\nu \rangle_{\rm amp}^{\MSbar} =
Z_\psi^{L,\MSbar}\,(Z^{L,\MSbar}_A)^{1/2}\sum_{i=1}^{13}\big((Z^{L,\MSbar})^{-1}\big)_{1i}  \langle\psi\,{\cal
  O}_i\,\bpsi\,A_\nu \rangle^{L}_{\rm amp}\,.
\label{GG3lattice}
\ee

The left-hand sides of the above equations are known from the calculations in DR,
see Subsection ~\ref{sec2.1}. The bare lattice Green's functions in these equations contain terms which 
diverge in the limit $a \to 0$; these divergent terms have a form similar to Eqs.~(\ref{G2}) and~(\ref{G3}), with two
differences:

\begin{itemize}
\item $\displaystyle \frac{1}{\epsilon} \rightarrow - \log(a^2)$
\item There are additional ${\cal O} \left(\displaystyle \frac{1}{a^2}\right)$, 
${\cal O}  \left(\displaystyle \frac{1}{a}\right)$ contributions:
\bea
\label{G2lattice}
{\rm in}\,\,\langle \psi\,{\cal O}_1\,\bpsi \rangle^{L}_{\rm amp}&:& \quad \rho_8\,(r_d\,\gamma_5\,\qqq + r_s\,\qq\,\gamma_5) + 
\rho_9\,i\,(r_d\,m_d + r_s\,m_s)\,\gamma_5 + \rho_{10}\cdot 1 \\[2ex]
{\rm in}\,\,\langle \psi\,{\cal O}_1\,\bpsi A_\nu \rangle^{L}_{{\rm amp}, 1PI}&:& \quad R_5\,g\,(r_d - r_s)\,\gamma_5\,\gamma_\nu\,.
\label{G3lattice}
\eea

\end{itemize}

These contributions lead to mixing with the lower dimension operators ${\cal O}_{11}$, 
${\cal O}_{12}$ and ${\cal O}_{13}$ defined in Eqs.~(\ref{O11}) - (\ref{O13}). 

The renormalization functions $Z^{L,\MSbar}_\psi$ ($Z^{L,\MSbar}_A$) for the quark (gluon) field, as well as $Z^{L,\MSbar}_g$, $Z^{L,\MSbar}_m$, were only partially available in the literature; we computed them for a general covariant gauge, using the Symanzik improved gauge action for different sets of values for the Symanzik coefficients. These results are presented in Appendix~\ref{app:Z}, in the RI$'$ renormalization scheme along with conversion factors to the $\MSbar$ scheme.

Renormalizability of the theory implies that the difference between the one-loop renormalized 
and bare Green's functions must only consist of expressions which are polynomial in $q_i,\, m$ ;
in this way, the right-hand sides of Eqs.~(\ref{GG2lattice}) - (\ref{GG3lattice}) can be rendered
equal to the corresponding left-hand sides, by an appropriate definition of the ($q_i$- and $m$-independent)
renormalization functions $Z_i^{L,\MSbar}$.
These differences can be written as follows:
\be
\label{RminusLattice2pt}
\langle \psi\,{\cal O}_1\,\bpsi \rangle_{\rm amp}^{\MSbar}-\langle \psi\,{\cal O}_1\,\bpsi \rangle_{\rm amp}^{L}=
g^2\Big(z_\psi^{L,\MSbar}-z_1^{L,\MSbar}\Big) \langle \psi \,{\cal O}_1\,\bpsi \rangle_{\rm tree} - \sum_{i=2}^{13}Z^{L,\MSbar}_i  \langle\psi\,{\cal O}_i\,\bpsi \rangle _{\rm tree}
\ee
and
\bea
\label{RminusLattice3pt}
\langle \psi\,{\cal O}_1\,\bpsi A_\nu \rangle_{\rm amp}^{\MSbar} - \langle \psi\,{\cal O}_1\,\bpsi A_\nu \rangle_{\rm amp}^{L}=&g^2&\Big(z_\psi^{L,\MSbar}+\frac{1}{2}z_A^{L,\MSbar}+ z_g^{L,\MSbar} - z_1^{L,\MSbar} \Big) \langle \psi \,{\cal O}_1\,\bpsi A_\nu \rangle _{\rm tree}\\\nonumber
\hspace*{1.5cm}&-& \sum_{i=2}^{13}Z^{L,\MSbar}_i \langle\psi\,{\cal O}_i\,\bpsi\,A_\nu \rangle _{\rm tree}\,.
\eea
Indeed, we have checked explicitly the polynomial character of Eqs.~(\ref{RminusLattice2pt}) - (\ref{RminusLattice3pt}).
This check is quite nontrivial, especially for Eq.~(\ref{RminusLattice3pt}), since both the bare and renormalized Green's functions, taken individually, exhibit a very complex dependence on the momenta $q_i$\,. The left-hand sides of Eqs.~(\ref{RminusLattice2pt}) - (\ref{RminusLattice3pt}) have the same tensorial form as Eqs.~(\ref{G2}) - (\ref{G3}), respectively, but with the additional contributions of Eqs.~(\ref{G2lattice}) - (\ref{G3lattice}).

Each tensorial structure (multiplying $\rho_1-\rho_{10},\,R_1-R_5$) will provide an equation; the set of
these equations (a total of 15) can be solved for the 13 mixing
coefficients $Z_i$. Two of the equations serve as consistency checks
and the remaining 13 lead to a well determined system. Upon solving all equations we obtain for the Iwasaki gluon action (see Appendix~\ref{app:mixCoef} for other gluon actions we have considered)
\bea
\label{Z1LMS}
Z^{L,\MSbar}_1  &=& 1+ \frac{g^2}{16\,\pi^2}\, \left[N_c \left(-7.9438 +\frac{1}{2}\,\log\left(a^2\,\bar\mu^2 \right) \right) + \frac{1}{N_c}  \left(4.4851  -\frac{5}{2}\,\log\left(a^2\,\bar\mu^2 \right) \right) \right] \\[1ex]             
Z^{L,\MSbar}_2  &=& \frac{g^2\,C_F}{16\,\pi^2}\,\left[4.5370 + 6\,\log\left(a^2\,\bar\mu^2 \right)\right] \\[1ex]
Z^{L,\MSbar}_3  &=& 0 \\[1ex]
\label{Z4LMS}
Z^{L,\MSbar}_4  &=& 0 \\[1ex]
Z^{L,\MSbar}_5  &=&  \frac{g^2}{16\,\pi^2}\,
                   \left[N_c \left(4.2758 -\frac{3}{2}\, \log\left(a^2\,\bar\mu^2 \right)\right) + 
                         \frac{1}{N_c}  \Bigl(-3.7777 + 3\,\log\left(a^2\,\bar\mu^2 \right)\Bigr) \right]\\[1ex]
Z^{L,\MSbar}_6  &=& 0 \\[1ex]
Z^{L,\MSbar}_7  &=& -\frac{Z^{L,\MSbar}_5}{2} \\[1ex]
Z^{L,\MSbar}_8  &=& \frac{g^2\,C_F}{16\,\pi^2}\,\left(-3.7760\right) \\[1ex]
Z^{L,\MSbar}_9  &=& \frac{Z^{L,\MSbar}_5}{2} \\[1ex]
Z^{L,\MSbar}_{10} &=& \frac{g^2\,C_F}{16\,\pi^2}\,\left[3.7777 - 3\,\log\left(a^2\,\bar\mu^2 \right)\right] \\[1ex]
Z^{L,\MSbar}_{11} &=& \frac{1}{a}\,\frac{g^2\,C_F}{16\,\pi^2}\,\left(-3.2020\right) \\[1ex]
\label{Z12LMS}
Z^{L,\MSbar}_{12} &=& -Z^{L,\MSbar}_{11} \\[1ex]
\label{Z13LMS}
Z^{L,\MSbar}_{13} &=& \frac{1}{a^2}\frac{g^2\,C_F}{16\,\pi^2}\,\left(36.0613\right)\,.
\eea
In these equations, $\bar{\mu}$ is the $\MSbar$ renormalization scale which appears in $\langle \psi\,{\cal O}_1\,\bpsi \rangle_{\rm amp}^{\MSbar}$ and $\langle \psi\,{\cal O}_1\,\bpsi A_\nu \rangle_{\rm amp}^{\MSbar}$ by virtue of: 
$g = \mu^{-\epsilon}\,(Z_g^{DR,\MSbar})^{-1}\,g_0^{DR},\, \bar\mu = \mu (4\pi/e^{\gamma_E})^{1/2}$.

The above results for $Z^{L,\MSbar}_1$ - $Z^{L,\MSbar}_{13}$ are independent of the choices $r_s=\pm 1$, $r_d=\pm 1$. 
There is also a small systematic error originating from the numerical estimation of lattice integrals, however it is much smaller than the displayed accuracy
of the results. 
It is important to emphasize that the coefficients $Z_{11}-Z_{13}$, which control the mixing with lower dimension operators, may receive also non-perturbative contributions proportional to $[(1/a) \exp(-1/(2 \beta_0 g_0^2))]^k \sim \Lambda^k$ (with $k=1,2$ for $Z_{11,12}$ and $Z_{13}$ respectively)~\cite{Maiani:1991az}.
For this reason, a proper subtraction of the mixing with operators ${\cal O}_{11} - {\cal O}_{13}$ must be implemented in a non-perturbative way~\cite{Constantinou:2014cra, Constantinou:2014wna,Constantinou:future}.

If one wants to renormalize in an (appropriately defined)
RI$'$ scheme, the calculation in DR is not
necessary: it suffices to compute the bare Green's functions on the
lattice. In this case the left-hand sides of
Eqs. (\ref{GG2}) - (\ref{GG3}), for particular values of the external
momenta, are dictated by the RI$'$ renormalization conditions.

The conversion factor between the RI$'$ and the $\overline{\rm MS}$
scheme will actually be a (13$\times$13) matrix in this case:
$C_{ij}^{RI',\MSbar}$. Since this matrix is regularization independent,
one may compute it through:
\be
{\cal O}_R^{\MSbar} \equiv C^{RI',\MSbar}
\,{\cal O}_R^{RI'}\,,\quad
C^{RI',\MSbar} = \left(Z^{DR,\MSbar}\right)^{-1}\,Z^{DR,RI'}.
\ee
Thus, in RI$'$, the mixing coefficients read (in matrix notation):
\be
Z^{L, RI'}= Z^{L,\MSbar} C^{RI',\MSbar}\,.
\ee

\section{Non-perturbative determination of the power-divergent mixing coefficients}
\label{Nonpert}

In this Section we present the non-perturbative determination of the coefficients $Z_{13}$ and $Z_{12}$ describing the power-divergent mixings of the chromomagnetic operator with the scalar and pseudoscalar densities, respectively.
We use lattice QCD simulations with the gauge configurations produced by ETMC with four flavors of dynamical quarks ($N_f = 2+1+1$), which include in the sea, besides two light mass degenerate quarks, also the strange and charm quarks with masses close to their physical values \cite{Baron:2010bv, Baron:2010th, Baron:2011sf}. 

Due to the equations of motion some of the operators ${\cal O}_1$ - ${\cal O}_{13}$ do not appear in the calculation of on-shell matrix elements. 
The remaining ones, namely ${\cal O}_1$, ${\cal O}_2$, ${\cal O}_3$,  ${\cal O}_4$,  ${\cal O}_{12}$ and  ${\cal O}_{13}$, are present and it is therefore crucial to have a reliable estimate of the corresponding mixing coefficients. 
For operators of the same dimensionality as the chromomagnetic one, i.e.~${\cal O}_1$ - ${\cal O}_4$, our perturbative one-loop results [see e.g.~Eqs.~(\ref{Z1LMS}-\ref{Z4LMS})] are expected to provide a satisfactory estimate.
However, as already discussed in Section \ref{Intro}, for the mixing coefficients of the operators with lower dimensionality ${\cal O}_{12}$ and ${\cal O}_{13}$, which are power-divergent, perturbation theory is expected to provide only a ballpark estimate \cite{Maiani:1992vl}.

In order to achieve a non-perturbative estimate of the mixing coefficients of the operators ${\cal O}_{12}$ and ${\cal O}_{13}$ we impose the following renormalization conditions \cite{Constantinou:2013zqa}
 \bea
      \label{c13}
      \lim_{m_s, ~ m_d \to 0} ~ \langle \pi | {\cal O}_1^{\rm R} | K \rangle & = & \frac{1}{Z_1} \lim_{m_s, ~ m_d \to 0} \langle \pi |{\cal{O}}_1 - 
                                                                                                                           \frac{c_{13}}{a^2}{\cal{O}}_{13} | K \rangle = 0 ~ , \\
     \label{c13&c12}
      \langle 0 |{\cal O}_1^{\rm R} | K \rangle_{m_s, m_d} & = & \frac{1}{Z_1} \langle 0 | {\cal{O}}_1 - \frac{c_{13}}{a^2}{\cal{O}}_{13} - 
                                                                                                   \frac{c_{12}}{a}{\cal{O}}_{12} | K \rangle_{m_s, m_d} = 0 ~ ,
 \eea
where the pion and kaon states are taken to be at rest\footnote{Such a choice is motivated by the fact that, as already observed, the matrix elements of the operator ${\cal{O}}_4 = \Box \left( \bpsi_s \psi_d \right)$ are proportional to the squared four-momentum transfer between the external hadronic states. This implies that for external $\pi$- and $K$-mesons at rest (or, more precisely, in the case of mesons with equal spatial momentum) the operator ${\cal{O}}_4$ has vanishing matrix elements in the SU(3) chiral limit and therefore it does not affect the determination of the mixing coefficient $c_{13}$ from Eq.~(\ref{c13}). The same happens for the contributions of the operators ${\cal{O}}_2$ and ${\cal{O}}_3$, since they are directly proportional to the quark masses.}.
Note that in this Section the operators ${\cal{O}}_i$ are the bare (local) lattice versions of the operators introduced in Section \ref{sec1} [see Eqs.~(\ref{O1}-\ref{O13}))]. 
In Eqs.~(\ref{c13}-\ref{c13&c12}) we have factorized out explicitly the power divergence of the mixings with the operators ${\cal{O}}_{12, 13}$ and we have introduced  the renormalization scale independent \cite{Testa:1998ez} mixing coefficients $c_{12,13}$, which can be written as appropriate combinations of operator renormalization constants (see Eq.~(\ref{ciZi}) below).
The renormalization conditions (\ref{c13}) and (\ref{c13&c12}) are valid up to cutoff effects, which are ${\cal{O}}(a^2)$ for Eq.~(\ref{c13}) and ${\cal{O}}(a)$ for Eq.~(\ref{c13&c12}).

The first condition (\ref{c13}) requires that the renormalized chromomagnetic operator ${\cal{O}}_1^{\rm R}$ has vanishing on-shell matrix elements in the SU(3) chiral limit.
This is also consistent with ChPT, which for the matrix element of ${\cal O}_1^{\rm R} \equiv 16 \pi^2 ~ Q_g^+$ predicts at LO [see Eqs.~(\ref{ME}-\ref{BCMO})] 
 \be
    \langle \pi^j(p_\pi) | {\cal{O}}_1^{\rm R} | K^j (p_K) \rangle = \frac{11}{2} C^j \frac{M_K^2}{m_s + m_d} (p_\pi \cdot p_K) B_{CMO} ~ ,
    \label{piQK}
\ee
where $C^\pm = 1$, $C^0 = -1 / \sqrt{2}$ and $B_{CMO}$ is a SU(3) ChPT low-energy constant.

The second condition (\ref{c13&c12}) imposes that in the continuum limit the parity violating matrix element $\langle 0 |{\cal{O}}_1^{\rm R} | K \rangle$ is identically vanishing. 

As usual, in the lattice version of the chromomagnetic operator (\ref{O1}) the gluon tensor $G_{\mu \nu}$ is replaced by its clover discretization $P_{\mu \nu}$, namely \cite{Gabrielli:1990us}
 \be
       \label{O1_a}
       {\cal {O}}_1 = g_0 ~ \overline{\psi}_s \sigma_{\mu \nu} P_{\mu \nu} \psi_d ~ , 
 \ee
where 
 \be 
    P_{\mu \nu}(x) \equiv \frac{1}{4 a^2} \sum_{j = 1}^4 \frac{1}{2 i g_0} \left[ P_j(x) - P_j^\dagger(x) \right]
    \label{Pmunu}
 \ee 
and the sum is over the four plaquettes $P_j(x)$ in the $\mu$-$\nu$ plane stemming from $x$ and taken in the counterclockwise sense.

Our lattice setup is the same as the one adopted in Ref.~\cite{Carrasco:2014cwa} for the determination of the up, down, strange and charm quark masses.
The fermions were simulated using the Wilson Twisted Mass Action \cite{Frezzotti:2000nk, Frezzotti:2003xj} which, at maximal twist, allows for automatic ${\cal{O}}(a)$-improvement \cite{Frezzotti:2003ni, Frezzotti:2004wz}.
In order to avoid the unphysical flavor mixing in the strange and charm valence sectors we adopted the non-unitary set up described in Ref.~\cite{Frezzotti:2004wz}, in which the valence quarks are regularized as OS fermions \cite{Osterwalder:1977pc}.
For the gauge links we simulated the Iwasaki action \cite{Iwasaki}, because it proved to facilitate simulations with light quark masses allowing to bring the simulated pion mass down to approximately $210$ MeV \cite{Baron:2010bv,Baron:2010th,Baron:2011sf}.

The details of the ETMC gauge ensembles are collected in Table \ref{tab:simudetails}, where the number of the gauge configurations analized ($N_{cfg}$) corresponds to a separation of $20$ trajectories.
At each lattice spacing, different values of the light sea quark mass were considered. 
The up and down quark masses were always taken to be degenerate and equal in the sea and valence sectors ($m_u = m_d = m_\ell$). 
The masses of both the strange and the charm sea quarks were fixed, at each $\beta$, to values close to the physical ones \cite{Baron:2010bv}.
We simulated quark masses for the light sector in the range $3 ~ m_\ell^{phys} \lesssim  m_\ell \lesssim 12 ~ m_\ell^{phys} $, while we used three values of the valence strange quark mass in the range $0.7 ~ m_s^{phys} \lesssim  m_s \lesssim 1.2 ~ m_s^{phys}$ in order to interpolate to the physical strange quark mass \cite{Carrasco:2014cwa}.
The simulated pion masses cover the range $\simeq 210 \div 450$ MeV.

\vspace{0.5cm}
\begin{table}[hbt!]
\begin{center}
\renewcommand{\arraystretch}{1.20}
\begin{tabular}{||c|c|c|c|c|c|c|c||}
\hline
ensemble & $\beta$ & $V / a^4$ & $a\mu_{sea} = a\mu_\ell$& $a\mu_\sigma$ & $a\mu_\delta$ & $N_{cfg}$ & $a\mu_s$ \\
\hline
$A30.32$ & $1.90$ & $32^{3}\times 64$ &$0.0030$ &$0.15$ &$0.19$ &$150$& $0.0145, 0.0185, 0.0225$ \\
$A40.32$ & & & $0.0040$ & & & $100$ & \\
$A50.32$ & & & $0.0050$ & & &  $150$ & \\
\cline{1-6} 
$A60.24$ & $1.90$ & $24^{3}\times 48 $ & $0.0060$ &$0.15$ & $0.19$& $150$ & \\
$A80.24$ & & & $0.0080$ & & &  $150$ & \\
$A100.24$ &   & & $0.0100$ & & &  $150$ & \\
\hline
$B25.32$ & $1.95$ & $32^{3}\times 64$ &$0.0025$&$0.135$ &$0.170$& $150$& $0.0141, 0.0180, 0.0219$ \\
$B35.32$ & & & $0.0035$  & & & $150$ & \\
$B55.32$ & & & $0.0055$  & & & $150$ & \\
$B75.32$ &  & & $0.0075$ & & & $80$ & \\
\cline{1-6}
$B85.24$ & $1.95$ & $24^{3}\times 48 $ & $0.0085$ &$0.135$ &$0.170$ & $150$ & \\
\hline
$D15.48$ & $2.10$ & $48^{3}\times 96$ &$0.0015$&$0.12$ &$0.1385 $& $60$& $0.0118, 0.0151, 0.0184$ \\ 
$D20.48$ & & & $0.0020$  &  &  & $100$ & \\
$D30.48$ & & & $0.0030$  &  &  & $100$ & \\
 \hline   
\end{tabular}
\renewcommand{\arraystretch}{1.0}
\end{center}
\normalsize
\caption{Details of the gauge ensembles and values of the simulated sea and valence quark bare masses used in this work (after Ref.~\cite{Carrasco:2014cwa}).}
\label{tab:simudetails}
\end{table}

Quark propagators with different valence masses are obtained using the multiple mass solver method \cite{Jegerlehner:1996pm, Jansen:2005kk}, which allows to invert the Dirac operator for several quark masses at a relatively low computational cost.   
The statistical accuracy of the meson correlators is significantly improved by using the ``one-end" stochastic method \cite{McNeile:2006bz}, which includes spatial stochastic sources at a single time slice chosen randomly.
Statistical errors are evaluated using the jackknife procedure.

In our lattice setup both the scalar and the pseudoscalar non-singlet densities renormalize multiplicatively and, choosing in particular $r_s = r_d = r$ (see below), we have
 \be
      \label{O13&O12}
      {\cal O}_{13}^{\rm R} = Z_P {\cal {O}}_{13}  = Z_P ~ \overline{\psi}_s \psi_d ~ , \qquad \qquad 
      {\cal O}_{12}^{\rm R} = \frac{Z_S}{Z_P}  {\cal {O}}_{12} = \frac{Z_S}{Z_P} ~ i r (\mu_s + \mu_d) \overline{\psi}_s \gamma_5 \psi_d ~ ,
 \ee
where $\mu = Z_P \cdot m$ is the (twisted) bare quark mass, while $Z_P$ and $Z_S$ are the renormalization constants of the (pseudo)scalar densities computed using the RI$^\prime$-MOM scheme in Ref.~\cite{Carrasco:2014cwa}.
Therefore, the mixing coefficients $c_{12}$ and $c_{13}$ introduced in Eqs.~(\ref{c13}-\ref{c13&c12}) are related to the operator renormalization constants via 
 \be
      \label{ciZi}
      \frac{c_{13}}{a^2} \equiv Z_{13} ~ Z_P ~ , \qquad \qquad
      \frac{c_{12}}{a} \equiv Z_{12} \frac{Z_S}{Z_P} ~ . 
 \ee 

In order to extract the coefficient $c_{13}$ from the condition (\ref{c13}) we have computed for each gauge ensemble the 3-point meson correlators defined as
 \be
      \label{C3j}
      C_3^j(t, t^\prime) = \frac{1}{L^6} \sum\limits_{\vec{x}, \vec{y}, \vec{z}} \left\langle 0 \right | P_5^\pi (y) {\cal{O}}_j (x) P_5^{K \dag} (z) \left| 0 \right\rangle 
                                      \delta_{t, (t_x  - t_z )} \delta_{t^\prime, (t_y  - t_z )} ~ ,
 \ee
where  $j = 1$ or $j = 13$, $P_5^\pi (x) = i \overline{\psi}_d(x) \gamma_5 \psi_u(x)$ and $P_5^K (x) = i \overline{\psi}_s(x) \gamma_5 \psi_u(x)$. 
Our choice $r_u = - r_d = - r_s$ guarantees that the two valence quarks in the pion and kaon mesons have always opposite values of the Wilson parameter.
In this way the pion and kaon states in the matrix elements of the renormalized chromomagnetic operator $\langle \pi | {\cal O}_1^{\rm R} | K \rangle$ have good scaling and chiral properties.

At large euclidean time the correlation function is dominated by the contribution of the lightest states and one gets
 \be
      C_3^j(t, t^\prime)_{ ~ \overrightarrow{t  \gg a, ~ (t^\prime - t) \gg a} ~ } \frac{Z_\pi}{2M_\pi} \frac{Z_K^*}{2M_K} 
              \langle \pi | {\cal{O}}_j | K \rangle ~ e^{ - M_\pi  t} ~ e^{ - M_K  (t^\prime - t)}
 \ee
with $Z_{\pi(K)} = \langle 0 | P_5^{\pi(K)} (0) | \pi(K) \rangle$.

From the large time behavior of the 3-point correlators corresponding to the chromomagnetic and scalar density insertions one can construct the following ratio 
 \be
      \label{R13}
      \frac{C_3^1(t, t^\prime)}{C_3^{13}(t, t^\prime)} _{ ~ \overrightarrow{t  \gg a, ~ (t^\prime - t) \gg a}  ~ } 
      \frac{ \langle \pi | {\cal{O}}_1 | K \rangle}{ \langle \pi | {\cal{O}}_{13} | K \rangle} \equiv ~ R_{13}(m_s, m_\ell; m_\ell),
 \ee
where the first two quark masses are those involved in the transition and the third one is the spectator valence quark mass, which is taken to be equal to the light sea quark mass.
According to Eq.~(\ref{c13}), the ratio $R_{13}(m_s, m_\ell; m_\ell)$ provides in the SU(3) chiral limit an estimate of the mixing coefficient $c_{13}$ at each value of the lattice spacing, namely
 \be
      \label{c13_a}
      c_{13} = \lim_{m_s, ~ m_\ell \to 0}  a^2 R_{13}(m_s, m_\ell; m_\ell) ~ .
 \ee
Note that the dimensionless quantity $a^2 R_{13}(m_s, m_\ell; m_\ell)$ is extracted directly from the ratio (\ref{R13}) computed in lattice units.

In order to perform the chiral limit we start by computing the ratio $C_3^1(t, t^\prime = T/2) / C_3^{13}(t, t^\prime = T/2)$ in the degenerate case $m_s = m_\ell$ for all the gauge ensembles of Table \ref{tab:simudetails}\footnote{Precisely, we replace in Eqs.~(\ref{O1_a})-(\ref{c13_a}) the strange quark $s$ by a light quark $d^\prime$ with mass $m_{d^\prime} = m_\ell$ and $r_{d^\prime} = r_d$. In this way we guarantee the absence of disconnected contributions and keep the discretization effects in the squared pion mass of order ${\cal O}(a^2 m_\ell)$.}, which will be referred to as the $\pi \to \pi$ channel.
In this channel the mixing with the operator ${\cal O}_4$ is absent and the mixing with the operator ${\cal O}_{12}$ is linear in the light quark mass, while the mixings with the operators ${\cal O}_2$ and ${\cal O}_3$ are quadratic in the quark mass.

\begin{figure}[htb!]
\centering
\includegraphics[scale=0.90]{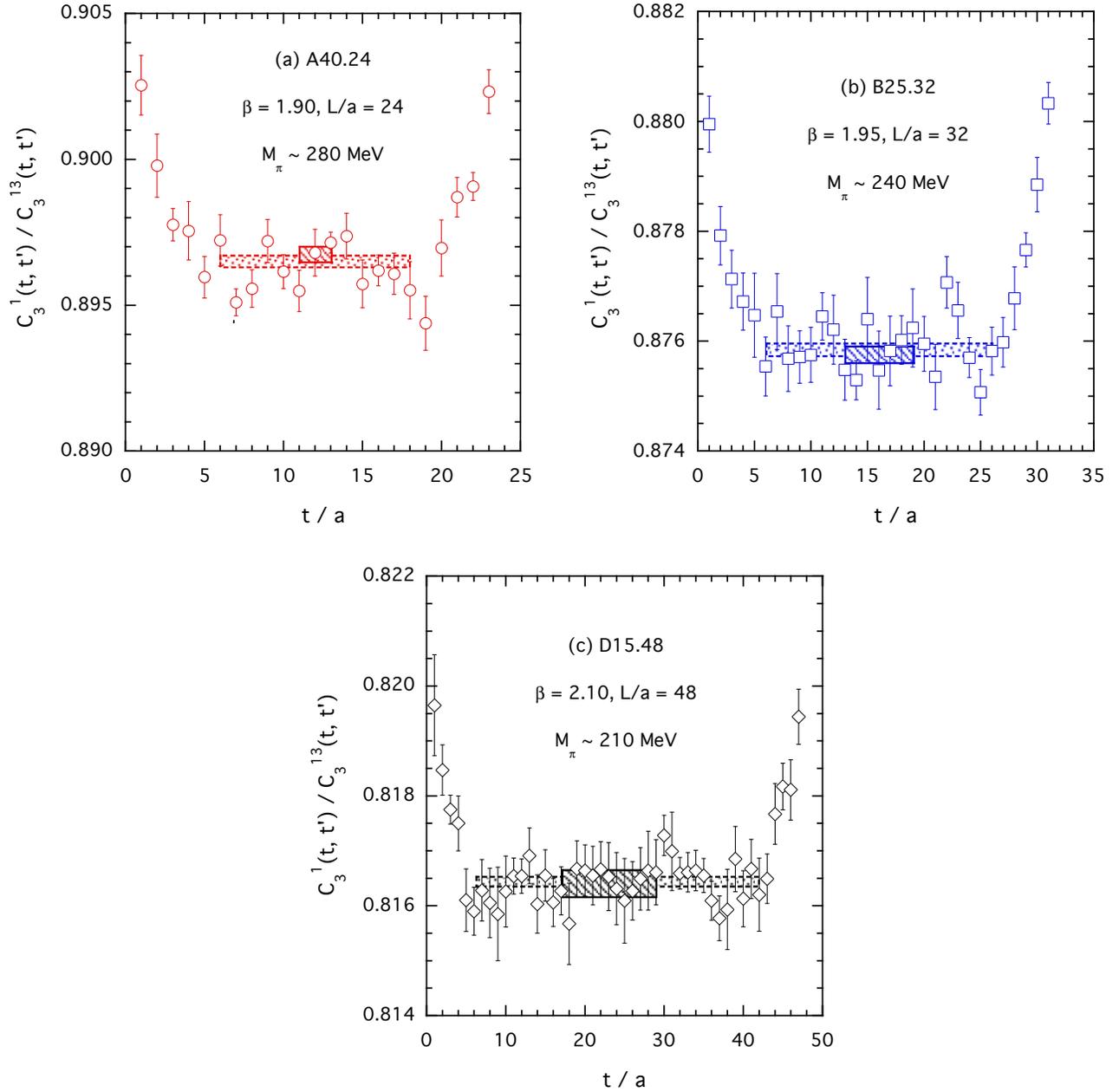}
\caption{The ratio of 3-point correlators $C_3^1(t, t^\prime) / C_3^{13}(t, t^\prime)$ in lattice units for the gauge ensembles A40.24 (a), B25.32 (b) and D15.48 (c) versus the insertion time $t$ for a fixed time separation $t^\prime = T / 2$ between the source and the sink. The horizontal solid (dashed) lines correspond to the central values and the errors obtained adopting the SP (LP) choice of the plateaux (see text).}
\label{fig_plateaux}
\end{figure}

The results obtained for the ensembles corresponding to the lightest pion mass at each of the three lattice spacings are presented in Fig.~\ref{fig_plateaux}.
It can be seen that the ratio $C_3^1(t, T/2) / C_3^{13}(t, T/2)$ exhibits nice plateaux.
In what follows we consider two different choices for the time intervals adopted to extract the values of $a^2 R_{13}(m_\ell, m_\ell; m_\ell)$ from the ratio $C_3^1(t, T/2) / C_3^{13}(t, T/2)$. 
The first choice, which will be referred to as the short plateaux (SP), corresponds to the time intervals $[t_{min}, T/2 - t_{min}]$, where $t_{min}$ is the time distance at which the pion ground state starts to dominate the corresponding 2-point correlator according to the analysis carried out  in Ref.~\cite{Carrasco:2014cwa} (see the horizontal solid lines in Fig.~\ref{fig_plateaux}). 
For the second choice, which will be referred to as the long plateaux (LP), the time intervals are extended from $6a$ to $T/2 - 6a$, as shown in Fig.~\ref{fig_plateaux} by the horizontal dashed lines.

\begin{figure}[htb!]
\centering
\includegraphics[scale=0.90]{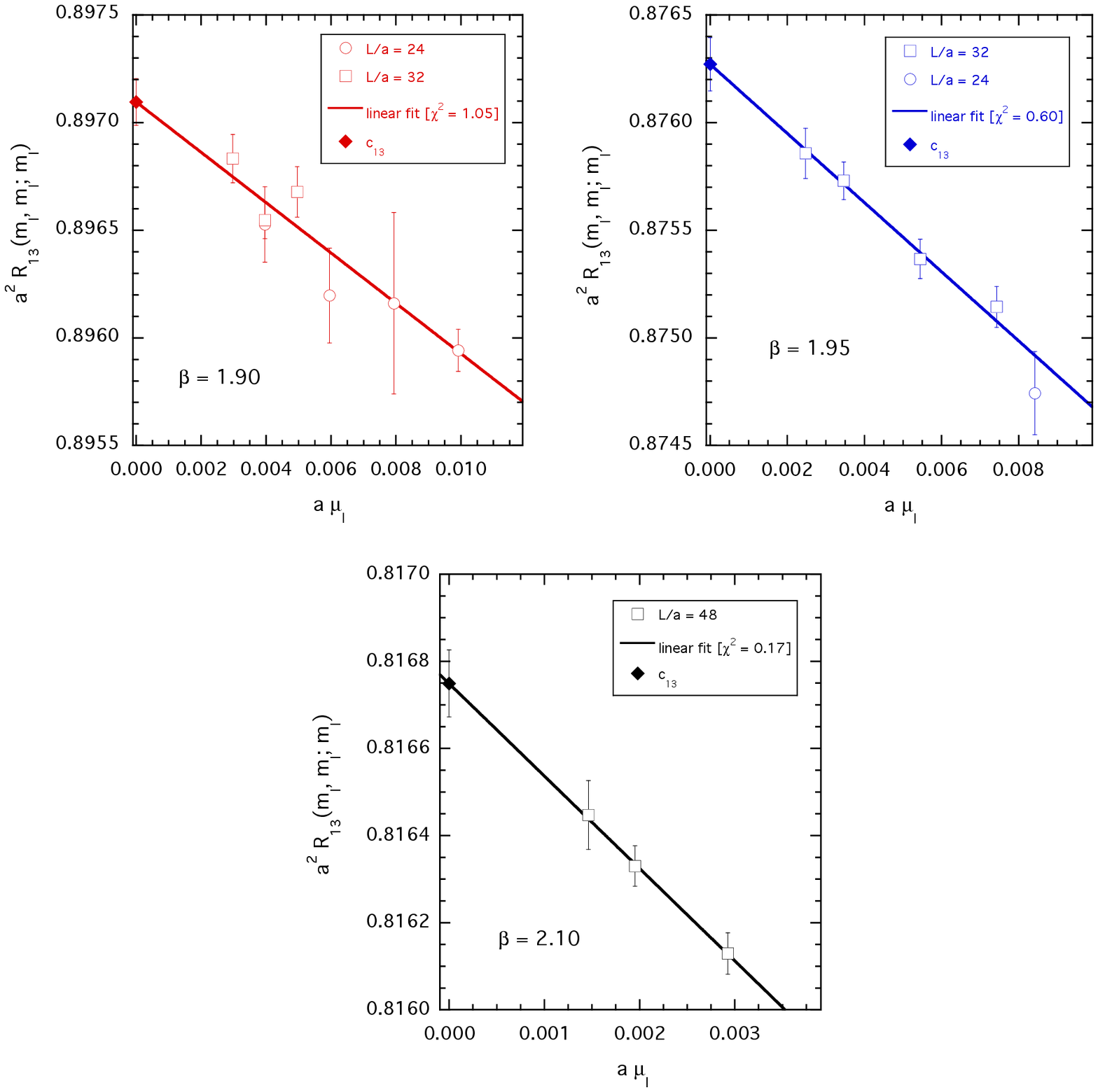}
\caption{The ratio $a^2 R_{13}(m_\ell, m_\ell; m_\ell)$, extracted using the LP choice for the plateaux, versus the (twisted) quark bare mass $a \mu_\ell = Z_P a m_\ell$ for each value of the lattice spacing. The solid lines are the results of linear fits in $a \mu_\ell$ applied to all data. The values of the $\chi^2$ variable (divided by the number of degrees of freedom) are reported in each inset. The diamonds represent the values of the mixing coefficient $c_{13}$ obtained in the chiral limit [see Eq.~(\ref{c13_a})].}
\label{fig_c13beta}
\end{figure}

The results obtained for $a^2 R_{13}(m_\ell, m_\ell; m_\ell)$, adopting the LP choice for the plateaux, are collected in Fig.~\ref{fig_c13beta} and show a linear dependence on the light quark mass, as expected at leading order in the quark mass expansion.
Therefore we fit the data at each lattice spacing using a linear ansatz of the form $a^2 R_{13}(m_\ell, m_\ell; m_\ell) = c_{13} + A \cdot a \mu_\ell$, where the parameters $c_{13}$ and $A$ are determined by minimizing the $\chi^2$ variable.
Similar results (with larger statistical errors) hold as well for the values of $a^2 R_{13}(m_\ell, m_\ell; m_\ell)$ obtained adopting the SP choice for the plateaux.

\begin{figure}[htb!]
\centering
\includegraphics[scale=0.90]{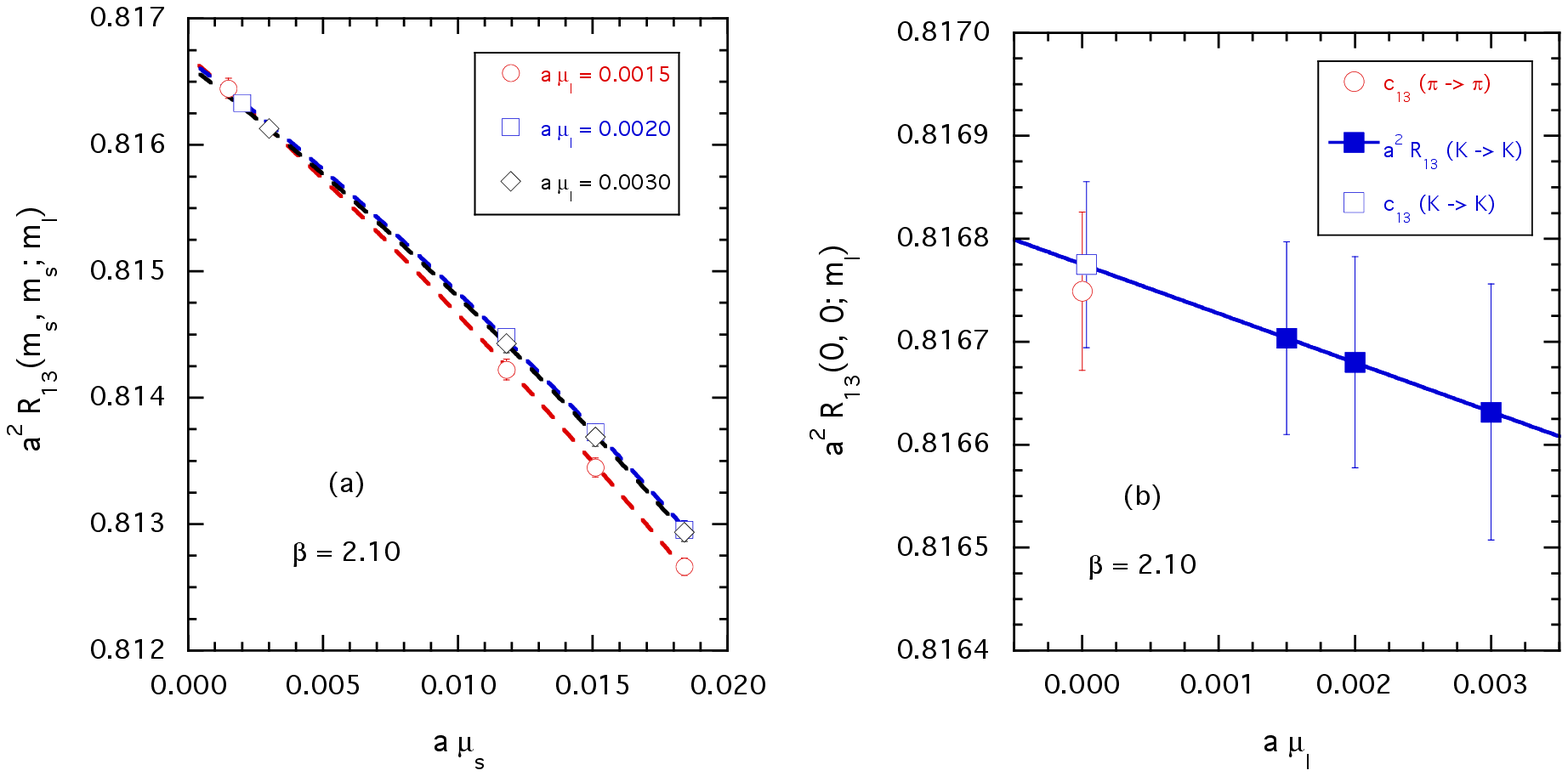}
\caption{(a) The ratio $a^2 R_{13}(m_s, m_s; m_\ell)$ in case of the $K \to K$ channel, adopting the LP choice for the plateaux, versus the strange quark bare mass $a \mu_s = Z_P a m_s$  for fixed values of the light quark mass $a \mu_\ell = Z_P a m_\ell$ corresponding to the three ensembles B15.48, B20.48 and B30.48 at $\beta = 2.10$. The dashed lines are quadratic fits in $a \mu_s$ applied to the data. (b) The ratio $a^2 R_{13}(0, 0; m_\ell)$ for the $K \to K$ channel versus the light-quark bare mass $a \mu_\ell$. The solid line is a linear fit in $a \mu_\ell$ and the empty square is the corresponding value of $c_{13}$. The empty circle is the result for $c_{13}$ obtained from the $\pi \to \pi$ channel [see Fig.~\ref{fig_c13beta} at $\beta = 2.10$].}
\label{fig_c13_210}
\end{figure}

As a check of the uncertainty related to the chiral extrapolation, we have also computed the ratio $a^2 R_{13}(m_s, m_s; m_\ell)$ using for $m_s$ the values corresponding to the valence strange (bare) quark mass reported in Table \ref{tab:simudetails}.
With respect to Eqs.~(\ref{R13})-(\ref{c13_a}) we replace the $d$ quark in the transition by a strange-like quark $s^\prime$ with mass $m_{s^\prime} = m_s$ and $r_{s^\prime} = r_s$. 
We refer to this channel as the $K \to K$ one. 
The results for the ratio $a^2 R_{13}(m_s, m_s; m_\ell)$ obtained for the ensembles at $\beta = 2.10$ are reported in Fig.~\ref{fig_c13_210}(a), where it can be seen that the data at fixed value of the light quark mass can be fitted adopting a quadratic ansatz in the strange quark mass.
In Fig.~\ref{fig_c13_210}(b) the SU(3) chiral point is finally reached by performing a linear fit in the light quark mass and the result is compared with the one corresponding to the $\pi \to \pi$ channel [see Fig.~\ref{fig_c13beta} at $\beta = 2.10$].
A good agreement between the two channels is obtained within one standard deviation.
Similar results are obtained at $\beta = 1.90$ and $1.95$.

Our determinations of the mixing coefficient $c_{13}$ obtained from the $\pi \to \pi$ and $K \to K$ matrix elements, adopting the two choices SP and LP for the plateaux, are presented in Table \ref{tab:c13}.
A very  good agreement is found between the results obtained in the two channels, while small differences (within at most $\approx 1.5$ standard deviations) are visible between the values corresponding to the SP and LP choices for the plateaux.
This comparison will be useful to quantify the uncertainties due to the subtraction of the mixing with the scalar density in the study of the (renormalized) chromomagnetic operator matrix elements \cite{Constantinou:2014tea, Constantinou:future}. 
Note that the uncertainties on $c_{13}$ are of the order of $0.01 \%$ in the case of the LP choice, while they do not exceed the level of $0.1 \%$ in the case of the SP choice.

\begin{table}[htb!]
\begin{center}
\renewcommand{\arraystretch}{1.20}
\begin{tabular}{||c||c|c||c|c||}
\hline
$\beta$ & \multicolumn{2}{|c||}{$\pi \to \pi$ channel} & \multicolumn{2}{|c||}{$K \to K$ channel} \\ \hline
             & SP & LP & SP & LP \\ \hline
$1.90$ & $0.89769 ~ (17)$ & $0.89710 ~ (11)$ & $0.89752 ~ (24)$ & $0.89716 ~ (10)$\\ \hline
$1.95$ & $0.87687 ~ (36)$ & $0.87627 ~ (12)$ & $0.87687 ~ (38)$ & $0.87632 ~ (13)$\\ \hline
$2.10$ & $0.81646 ~ (78)$ & $0.81675 ~ (08)$ & $0.81635 ~ (61)$ & $0.81677 ~ (08)$\\ \hline
\end{tabular}
\renewcommand{\arraystretch}{1.0}
\end{center}
\normalsize
\caption{Values of the mixing coefficient $c_{13}$ obtained as the chiral limit of the data on the ratios $a^2 R_{13}(m_\ell, m_\ell; m_\ell)$ for the $\pi \to \pi$ channel and $a^2 R_{13}(m_s, m_s; m_\ell)$ for the $K \to K$ channel, extracted at three values of the lattice spacing using the SP and LP choices for the plateaux.}
\label{tab:c13}
\end{table}

We now want to compare our non-perturbative results of Table \ref{tab:c13} with the predictions of perturbation theory at one loop obtained in the previous Section.
Since the mixing coefficient $Z_{13}$ starts already at order ${\cal O}(g^2)$, while both $Z_1$ and $Z_P$ start at order ${\cal O}(1)$, the one-loop perturbative term for $c_{13}$, defined in Eq.~(\ref{ciZi}), is the same as the one of $a^2 Z_{13}$, namely in the case of the ETMC action (see Eq.~(\ref{Z13LMS}))
 \be
      \label{c13PT}
      c_{13}^{\rm 1-loop ~ PT} = \frac{g^2 C_F}{16 \pi^2} ~ 36.0613 ~ .
 \ee 

\begin{figure}[htb!]
\centering
\includegraphics[scale=0.60]{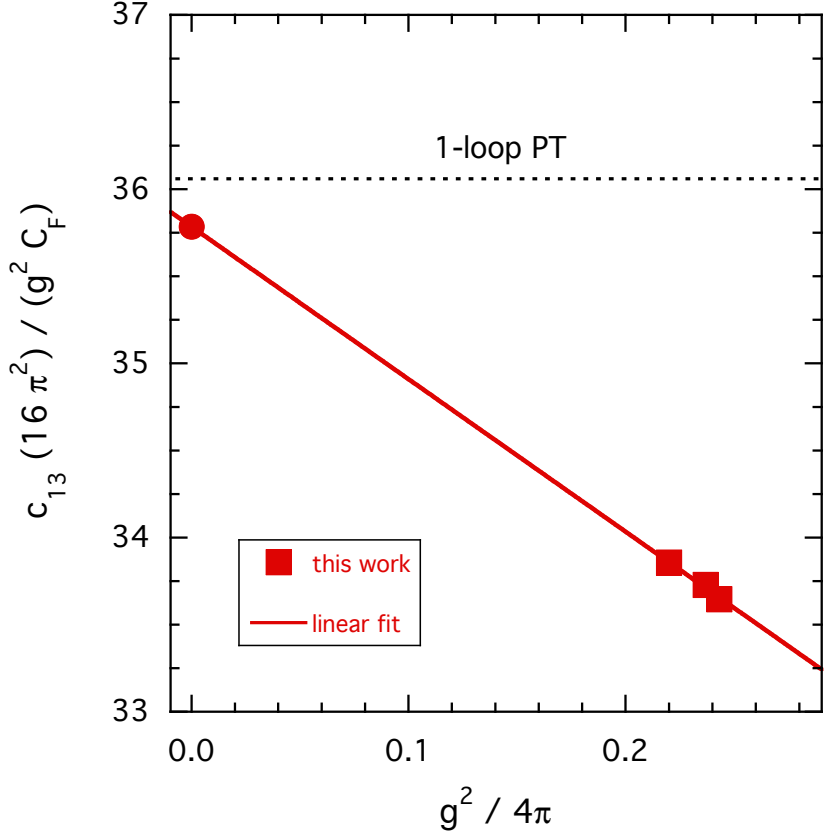}
\caption{The quantity $c_{13} (16 \pi^2) / (g^2 C_F)$ versus the coupling $g^2 / (4 \pi) = g_0^2 / (4 \pi) = 3  / (2 \pi \beta)$ calculated non-perturbatively at three values of the lattice spacings using the LP choice for the plateaux in the $\pi \to \pi$ channel (see Table \ref{tab:c13}). The horizontal dotted line is the prediction of lattice perturbation theory at one loop corresponding to Eq.~(\ref{c13PT}). The solid line corresponds to a linear fit in $g^2 / (4 \pi)$ and the full circle is the corresponding extrapolated value at $g^2 = 0$.}
\label{fig_c13PT}
\end{figure}

In Fig.~\ref{fig_c13PT}, using $g^2 = g_0^2 = 6 / \beta$, the non-perturbative results for $c_{13} (16 \pi^2) / (g^2 C_F)$, obtained at three values of the lattice spacing using the LP choice for the plateaux, are compared with the corresponding perturbative result from Eq.~(\ref{c13PT}).
It can be seen that the non-perturbative determinations of $c_{13}$ differ less than $\simeq 10 \%$ from the perturbative predictions at one loop at all values of the lattice spacing.
We expect however that in order to get a reliable determination of the renormalized CMO matrix elements a high-precision determination of $c_{13}$, at the level of $0.1 \%$ or better, will be required \cite{Constantinou:2014cra, Constantinou:2014wna}. 

Let us now discuss the non-perturbative determination of the mixing coefficient $c_{12}$ of the chromomagnetic operator with the pseudoscalar density.
To this end we make use of Eq.~(\ref{c13&c12}) and of our accurate non-perturbative results for the mixing coefficient $c_{13}$.
We anticipate that the ${\cal{O}}(a)$ terms affecting the r.h.s.~of Eq.~(\ref{c13&c12}), which are unavoidably relevant in the parity violating matrix elements $\langle 0 | {\cal{O}}_1 | K \rangle$ and $\langle 0 | {\cal{O}}_{13} | K \rangle$, turn out to be numerically competitive for our values of the lattice spacing with the contribution of the power-divergent mixing with the pseudoscalar density ${\cal{O}}_{12}$.
This finding can be expected also from the smallness of the one-loop perturbative estimate of $c_{12}$ (see Eq.~(\ref{Z12LMS})) with respect to the corresponding result (\ref{Z13LMS}) for the mixing coefficient $c_{13}$.

We start by computing for each gauge ensemble the 2-point meson correlators defined as
 \bea
      \label{C2PS}
      C_2^K(t) & = & \frac{1}{L^3} \sum\limits_{\vec{x}, \vec{z}} \left\langle 0 \right| P_5^K(x) P_5^{K \dag}(z) \left| 0 \right\rangle 
                               \delta_{t, (t_x  - t_z )} ~ , \\
      \label{C2j}
      C_2^j(t) & = & \frac{1}{L^3} \sum\limits_{\vec{x}, \vec{z}} \left\langle 0 \right| {\cal{O}}_j (x) P_5^{K \dag}(z) \left| 0 \right\rangle \delta_{t, (t_x  - t_z )} ~ , 
  \eea
where  $j = 1$ or $j = 13$ and $P_5^K (x) = i \overline{\psi}_s(x) \gamma_5 \psi_d(x)$ with $r_s = r_d$ (see Eq.~(\ref{O13&O12})).
 At large time distances one has
 \bea
       C_2^K(t) & _{ ~ \overrightarrow{t  \gg a} ~ } & \frac{|Z_K|^2}{2 M_K} e^{-M_K t} ~ , \nonumber \\
       C_2^j(t) & _{ ~ \overrightarrow{t  \gg a} ~ } & \frac{Z_K \langle 0 | {\cal{O}}_j | K \rangle}{2 M_K} e^{-M_K t} 
       \label{C2PS_larget}       
\eea
with $Z_K = \langle 0 | P_5^K | K \rangle$. 
We stress that, since $r_s = r_d$, the two valence quarks in the kaon have the same value of the Wilson parameter and therefore the squared meson mass $M_K^2$ differs from its continuum counterpart by terms of order ${\cal O}(a^2)$, which do not vanish in the chiral limit.

Then, using our non-perturbative results for $c_{13}$, the one-loop perturbative estimates for $Z_1$ (see Eq.~(\ref{Z1LMS})) and the non-perturbative renormalization constant $Z_S$ from Ref.~\cite{Carrasco:2014cwa}, we have computed the following ratio 
 \be
      \label{CMO_R12}
      R_{12}(m_\ell, m_s; a)  \equiv  \frac{1}{Z_1 Z_S \langle 0 | P_5^K | K \rangle} \left[ \langle 0 | {\cal{O}}_1 | K \rangle - \frac{c_{13}}{a^2} 
                                                        \langle 0 | {\cal{O}}_{13} | K \rangle \right] 
 \ee
both in the degenerate case $m_s = m_\ell$ and for the three simulated values of $m_s$ reported in Table \ref{tab:simudetails}. 
For the dimensionless quantity $a r_0 R_{12}(m_s, m_\ell; a)$, where $r_0$ is the Sommer parameter, the continuum limit expectation is a simple linear dependence on the sum of the light and strange (renormalized) quark masses, viz.
 \be
    \label{R12_naive}
     a r_0 R_{12}(m_\ell, m_s; a) = c_{12} \frac{Z_P}{Z_S Z_1} r_0 (m_\ell + m_s) + {\cal{O}}(a^2) ~ ,
 \ee
where the factor $Z_P / (Z_S Z_1)$ is almost constant for our three $\beta$-values, namely $Z_P / (Z_S Z_1) = \{1.285,1.275, 1.260\}$ for $\beta = \{1.90, 1.95, 2.10 \}$.
Note that the ratio $Z_P / Z_S$ is both scheme and renormalization scale independent, while the renormalization constant $Z_1$ carries the scheme and renormalization scale dependence of the l.h.s.~of Eq.~(\ref{R12_naive}).
Thus the coefficient $c_{12}$ is both scheme and renormalization scale independent (see Ref.~\cite{Testa:1998ez}).

However, as anticipated, the contribution of the lattice artefacts in Eq.~(\ref{R12_naive}) is not negligible. 
Beyond trivial terms proportional to $a^2$ and $a^2 (m_\ell + m_s)$, coming from the mixing of ${\cal{O}}_1$ with dimension-6 parity-odd operators, one should also consider that in the twisted-mass approach the Symanzik expansion of correlators like $C_2^j(t)$ (see Eq.~(\ref{C2j})) may contain chirally enhanced cutoff effects that are described by the (space-time integrated) insertion of the parity-odd local operator ${\cal{L}}_{\rm odd} = {\cal{L}}_5 + {\cal{O}}(a^3)$, connecting the vacuum with the one-pion state \cite{Frezzotti:2005gi, Dimopoulos:2009qv}.
In particular the matrix element of the chromomagnetic operator $\langle 0 | {\cal O}_1^R | K \rangle$ may receive a cutoff effect proportional to $\langle 0 | {\cal{L}}_5 |\pi \rangle \langle \pi | {\cal O}_1^R | K \rangle / M_\pi^2$.
Therefore, taking into account that at maximal twist we have $\langle 0 | {\cal{L}}_5 | \pi \rangle \propto a m_\ell$  \cite{Frezzotti:2005gi}, the following fit has been performed
 \bea
     a r_0 R_{12}(m_s, m_\ell; a) & = & \left(c_{12} \frac{Z_P}{Z_S Z_1} + d_{12} a^2 \right) r_0 (m_\ell + m_s) + h_0 ~ a^2 + h_0^\prime ~ a^4 \nonumber \\
                                                   & + & h_C ~ a^2 r_0 m_\ell \frac{M_K}{M_\pi} + h_S ~ a^2 r_0 m_\ell \frac{(M_K - M_\pi)^2}{M_\pi^2}  ~ ,
     \label{R12_fit}
 \eea 
where $c_{12}$, $d_{12}$, $h_0$, $h_0^\prime$, $h_C$ and $h_S$ are fitting parameters and the last two terms come from the expected mass dependence of the matrix elements of the chromomagnetic operator, $\langle \pi | {\cal{O}}_1^R | K \rangle \propto M_\pi M_K$, and of the operator ${\cal{O}}_4 = \Box \left( \bpsi_s \psi_d \right)$, $\langle \pi | \Box \left( \bpsi_s \psi_d \right) | K \rangle \propto (M_K - M_\pi)^2$.

The kaon mass $M_K$ appearing in Eq.~(\ref{R12_fit}) is directly extracted from the lattice correlator $C_2^K(t)$ (see Eq.~(\ref{C2PS})) and contains ${\cal O}(a^2)$ terms, which make it heavier than its continuum counterpart.
Instead the pion mass $M_\pi$ appearing in Eq.~(\ref{R12_fit}) may receive a tower of cutoff effects by multiple insertions of the parity-odd local operator ${\cal{L}}_{\rm odd}$, so that it is not known a priori whether it may be lighter or heavier than its continuum counterpart. 
Therefore we have adopted for the pion mass two choices differing by ${\cal{O}}(a^2)$ terms, namely the $OS$ pion $M_\pi^{OS}$ \cite{Osterwalder:1977pc}, which coincides with $M_K$ in the mass-degenerate case $m_s = m_\ell$, and the twisted-mass neutral pion $M_{\pi^0}^{TM}$, which for our lattice setup turns out be lighter than the charged one.
For the values of $M_{\pi^0}^{TM}$ we have used the lattice results reported in Ref.~\cite{Herdoiza:2013sla}. 

We have applied the fitting function (\ref{R12_fit}) to the lattice data for $a r_0 R_{12}(m_s, m_\ell; a)$ and found that the use of $M_{\pi^0}^{TM}$ produces smaller values of $\chi^2$ with respect to the OS choice $M_\pi^{OS}$.
Moreover we have tried also to insert an extra cutoff effect in the pion mass of the form $M_\pi^2 = [M_\pi^{TM}]^2 + c a^2$. 
However, the resulting value of the fitting parameter $c$ turned out to be largely compatible with $c = 0$ and no improvement in the $\chi^2$ value was found.

\begin{figure}[htb!]
\centering
\includegraphics[scale=0.90]{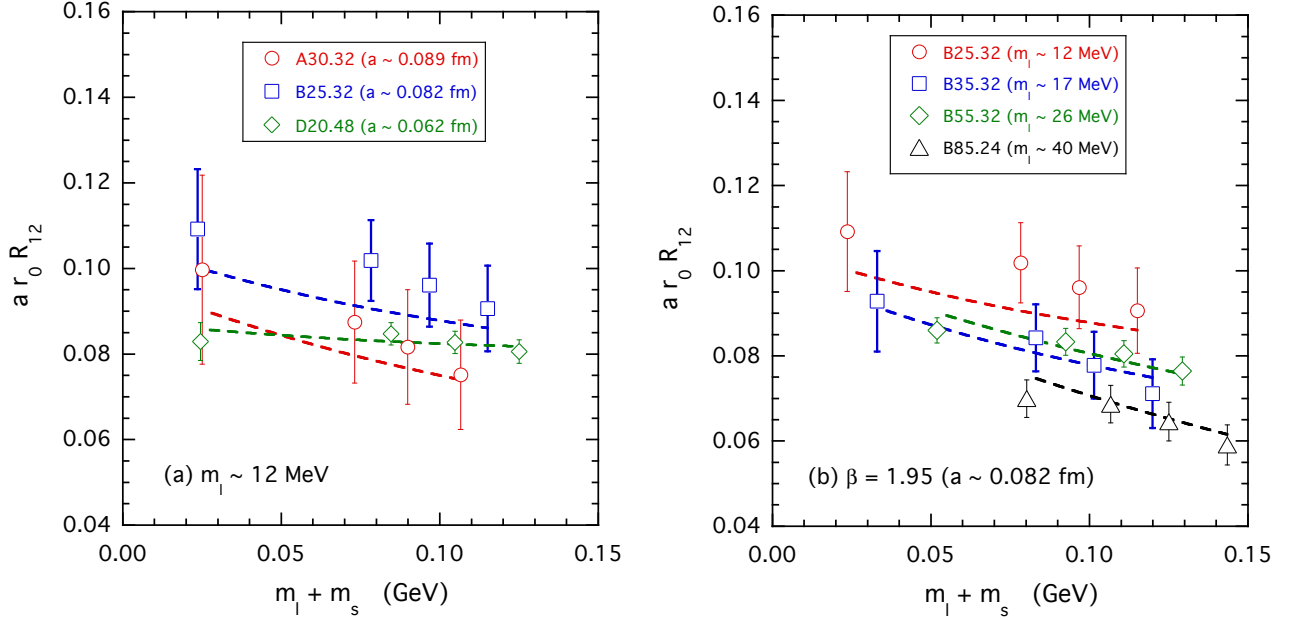}
\caption{The dimensionless quantity $a r_0 R_{12}(m_\ell, m_s; a)$ versus the (renormalized) quark mass ($m_\ell + m_s$), calculated using the values of $r_0 / a$ from Ref.~\cite{Carrasco:2014cwa} and the non-perturbative result for $c_{13}$ corresponding to the $\pi \to \pi$ channel and to the LP choice for the plateaux (see Table \ref{tab:c13}). In (a) the dots, squares and diamonds correspond to the results obtained for the gauge ensembles A30.32, B25.32 and D20.48, respectively, which share a (renormalized) light-quark mass $m_\ell \approx 12$ MeV. In (b) the results corresponding to the ETMC ensembles with $\beta = 1.95$ (fixed value of the lattice spacing) and various values of the light-quark mass $m_\ell$ (see the inset) are presented. The dashed lines represent the best fit curves according to Eq.~(\ref{R12_fit}).}
\label{fig_R12}
\end{figure}

The good quality of our best fit (corresponding to $\chi^2/d.o.f.~\simeq 0.6$) is shown in Fig.~\ref{fig_R12}.
In particular it can be seen that Eq.~(\ref{R12_fit}) is able to take properly into account the dependence on both the lattice spacing (see panel (a)) and the light-quark mass (see panel (b)).

The final non-perturbative result for the mixing coefficient $c_{12}$ is
 \be
       c_{12} = 0.035 ~ (20) ~ ,
       \label{c12_NP}
 \ee
which is approximately $40 \%$ of the value expected from one-loop perturbation theory for the ETMC action, namely (see Eq.~(\ref{Z12LMS}))
 \be
      \label{c12_PT}
      c_{12}^{\rm 1-loop ~ PT} = \frac{g^2 C_F}{16 \pi^2} ~ 3.2020 = \{ 0.0854, 0.0832, 0.0772\} \qquad {\rm for} \qquad \beta = \{1.90, 1.95, 2.10\} ~ .
 \ee 

The uncertainty in the non-perturbative determination (\ref{c12_NP}) of $c_{12}$ is $\approx 60 \%$, and this justifies {\it a posteriori} that the $g^2$-dependence of the mixing coefficient $c_{12}$ has been neglected in the present analysis.

We note, in conclusion, that the smallness of both the perturbative (\ref{c12_PT}) and non-perturbative (\ref{c12_NP}) determinations for $c_{12}$ implies that in our lattice formulation the subtraction of the power-divergent mixing with the pseudoscalar density is not going to play a crucial role for the numerical determination of the CMO matrix elements.

\section{Summary}
\label{summary}

The study of the chromomagnetic operator on the lattice has been hampered up to now by the complicated pattern of operator mixing. 
We identified these operators based on the symmetries of the regularized theory, which has been conveniently chosen to be twisted-mass LQCD (at maximal twist) with the Iwasaki gluon action.  

There are mixings with lower dimensional operators (which are power divergent), as well as with gauge non-invariant operators. 
We have computed all relevant mixing coefficients to one loop in lattice perturbation theory; this has required the calculation of both 2-point (quark-quark) and 3-point (quark-quark-gluon) Green's functions at nonzero quark masses. We have calculated all the elements of the mixing matrix that is relevant for the renormalization of the chromomagnetic operator at one loop in lattice perturbation theory.

For the first time the $1 / a^2$-divergent mixing of the chromomagnetic operator with the scalar density has been determined non-perturbatively with high precision (see Table \ref{tab:c13}). 
The $1 / a$-divergent mixing with the pseudoscalar density, which is peculiar of the twisted-mass formulation, has been also calculated non-perturbatively (see Eq.~(\ref{c12_NP})) and found to be smaller than its one-loop perturbative estimate (\ref{c12_PT}).
We have carried out the QCD simulations on the lattice using the gauge configurations produced by ETMC with $N_f = 2 + 1 + 1$ dynamical quarks, which include in the sea, besides two light mass degenerate quarks, also the strange and charm quarks with masses close to their physical values. 
Three values of the lattice spacing between $\simeq 0.6$ and $\simeq 0.9$ fm and pion masses in the range $210 \div 450$ MeV have been considered.

The results presented in this paper, which determine the mixing pattern of the chromomagnetic operator, are an essential ingredient for the determination of the (renormalized) CMO matrix element between pion and kaon states, whose calculation is in progress.
Preliminary results have been presented in Ref.~\cite{Constantinou:2014tea} and the final ones will be the subject of a forthcoming publication~\cite{Constantinou:future}.

\section*{Acknowledgments} 

We warmly thank G.C.~Rossi for valuable discussions about the non-perturbative determination of the $c_{12}$ mixing coefficient.

We acknowledge the CPU time provided by the PRACE Research Infrastructure under the project PRA027 ``QCD Simulations for Flavor Physics in the Standard Model and Beyond'' on the JUGENE BG/P system at the J\"ulich SuperComputing Center (Germany), and by the agreement between INFN and CINECA under the specific initiative INFN-lqcd123 on the Fermi system at CINECA (Italy).

M.~Constantinou and M.~Costa acknowledge financial support from the Cyprus Research Promotion Foundation under contract number TECHNOLOGY/$\Theta$E$\Pi$I$\Sigma$/0308(BE)/17. 

V.~L., G.~M.~and S.~S.~thank MIUR (Italy) for partial support under Contract No. PRIN 2010-2011 (DAMESYFLA).

D.~M.~acknowledges MIUR (Italy) for financial support under the program Futuro in Ricerca 2010 (RBFR10O36O).

\appendix

\section{Mixing coefficients $Z_i$}
\label{app:mixCoef}

In this Appendix we present our results for the mixing coefficients, $Z_i$ ($i=1,\ldots,13$) in the $\MSbar$ scheme,  for the following gluon actions: Wilson, tree-level Symanzik, Tadpole Improved L\"{u}scher-Weisz (TILW, at $\beta\,c_0=8.30$; $\displaystyle\beta=2N_c/g^2$), Iwasaki and Doubly Blocked Wilson (DBW2). The values of the Symanzik coefficients corresponding to these actions are collected in Table \ref{tb:SymCoef}.

\begin{table}[htb!]
\begin{center}
\begin{tabular}{| c| c| c| c| c| c|}
\hline
\,\,Coefficient\,\, & \,\,Wilson\,\, &\,\, Tree-level Symanzik \,\,& \,\,TILW ($\beta\,c_0=8.30$)\,\, & \,\,Iwasaki\,\, & \,\,DBW2\,\,\\\hline
$c_0$&1&5/3&2.386978&3.648&12.2688\\\hline
$c_1$&0&-1/12&-0.159128&-0.331&-1.4086\\\hline
$c_2$&0&0&0&0&0\\\hline 
$c_3$&0&0&-0.014244&0&0\\\hline
\end{tabular}
\caption{Symanzik coefficients for various choices of gluon actions.}
\label{tb:SymCoef}
\end{center}
\end{table}

Our calculation has been performed in an arbitrary covariant gauge. 
All the mixing coefficients $Z_i$ ($i = 1, \ldots,13$) in the $\MSbar$ scheme are gauge independent.
The generic forms of the mixing coefficients are
\bea
Z^{L,\MSbar}_1  &=& 1+ \frac{g^2}{16\,\pi^2}\,\left[N_c \left(e_{1,1} +\frac{1}{2}\,\log(a^2\,\bar\mu^2) \right) + \frac{1}{N_c}\left(e_{1,2} -\frac{5}{2}\,\log(a^2\,\bar\mu^2) \right)\right]\,,\\
Z^{L,\MSbar}_2  &=& \frac{g^2\,C_F}{16\,\pi^2}\,\left[e_{2} + 6\,\log\left(a^2\,\bar\mu^2 \right)\right]\,,\\
Z^{L,\MSbar}_3  &=& 0\,,\\
Z^{L,\MSbar}_4  &=& 0\,,\\
Z^{L,\MSbar}_5  &=& \frac{g^2}{16\,\pi^2}\,\left[N_c \left(e_{5,1} -\frac{3}{2}\,\log(a^2\,\bar\mu^2) \right) + \frac{1}{N_c}\left(e_{5,2} +3\,\log(a^2\,\bar\mu^2) \right) \right]\,,\\ 
Z^{L,\MSbar}_6  &=& 0\,,\\
Z^{L,\MSbar}_7  &=& -\frac{Z^{L,\MSbar}_5}{2}\,,\\
Z^{L,\MSbar}_8  &=& \frac{g^2\,C_F}{16\,\pi^2}\,\left(e_{8}\right)\,,\\
Z^{L,\MSbar}_9  &=& \frac{Z^{L,\MSbar}_5}{2}\,,\\
Z^{L,\MSbar}_{10} &=& \frac{g^2\,C_F}{16\,\pi^2}\,\left[-e_{5,2} - 3\,\log\left(a^2\,\bar\mu^2 \right)\right]\,,\\
Z^{L,\MSbar}_{11} &=& \frac{1}{a}\,\frac{g^2\,C_F}{16\,\pi^2}\,\left(e_{11}\right)\,,\\
Z^{L,\MSbar}_{12} &=& -Z^{L,\MSbar}_{11}\,,\\
Z^{L,\MSbar}_{13} &=& \frac{1}{a^2}\frac{g^2\,C_F}{16\,\pi^2}\,\left(e_{13}\right)\,,
\eea
where the values of $e_{i}$, $e_{i,j}$ are shown explicitly in Table~\ref{tb:MixCoef}.

\begin{table}[htb!]
\begin{center}
\begin{tabular}{| c| c| c| c| c| c|}
\hline
\,\,Coefficient\,\, & \,\,Wilson\,\, & \,\,Tree-level Symanzik\,\, & \,\,TILW ($\beta\,c_0=8.30$)\,\, & \,\,Iwasaki\,\, &\,\, DBW2\,\,\\\hline

$e_{1,1}$&-16.8770&-12.8455&-10.4920&-7.9438&-3.2465\\\hline

$e_{1,2}$&13.4540&9.3779&7.0022&4.4851&-0.5102\\\hline

$e_{2}$&1.9290&2.7677&3.4589&4.5370& 8.5250\\\hline

$e_{5,1}$&5.9806&5.3894&4.9311&4.2758&2.2834\\\hline

$e_{5,2}$&-6.4047&-5.5061& -4.8014&-3.7777&-0.5292\\\hline

$e_{8}$&-4.0626&-3.9654&-3.8894&-3.7760&-3.4713\\\hline

$e_{11}$&-4.4977&-4.0309&-3.6792&-3.2020&-1.9216\\\hline

$e_{13}$&54.9325&47.7929&42.6253&36.0613&19.9812\\\hline
\end{tabular}
\caption{Results for the mixing coefficients at one-loop using the $\MSbar$ scheme on the lattice. 
The finite parts $e_{i}$ and $e_{i,j}$ are given for five actions: Wilson, tree-level Symanzik, TILW ($\beta\,c_0=8.30$), Iwasaki and DBW2.}
\label{tb:MixCoef}
\end{center}
\end{table}

\section{Perturbative one-loop Renormalization of $Z_c$\,, $Z_\psi$\,, $Z_m$\,, $Z_A$\,, $Z_g$ on the Lattice}
\label{app:Z}

In this Appendix we provide the results of our one-loop calculation for the renormalization functions of the ghost field ($Z_c$), quark field ($Z_\psi$), gluon field ($Z_A$), coupling constant ($Z_g$), quark mass ($Z_m$). These functions enter the renormalization of the chromomagnetic operator through Eqs.~(\ref{ZgRules2b}),~(\ref{GG2}),~(\ref{GG3}),~(\ref{RminusLattice3pt}),~(\ref{Otilde2}). The computation was performed using twisted mass fermions, Symanzik improved gluon action and a general covariant gauge. Here we present the results for Wilson, tree-level Symanzik, TILW ($\beta\,c_0=8.30$), Iwasaki and DBW2 actions.
For the extraction of the renormalization functions, we applied the RI$'$ scheme at a scale $\bar{\mu}$. Once we have computed the renormalization functions in the RI$'$ scheme we can construct their $\MSbar$ counterparts using conversion factors which are known (see, e.g.,~\cite{Gracey:2003yr}), up to the required perturbative order. 

The aforementioned renormalization functions are defined as follows
\bea
g_0 &=& Z_g\,g, \\
c &=& \sqrt{Z_c}\,c^R, \\
\psi &=& \sqrt{Z_\psi}\,\psi^R,\\
A_{\mu} &=& \sqrt{Z_A}\,A^R_{\mu},\\ \label{B5}
\alpha &=& Z_\alpha^{-1} Z_A\,\,\alpha^R,\\
m &=& Z_m m^R\,.
\eea
In the above, $Z_g$ actually stands for $Z^{L,RI'}_g$; similarly for all other Z's.
The renormalization function $Z_\alpha$ for the gauge parameter receives no one-loop contribution.

\subsection{Ghost Field Renormalization $Z_c$}

\begin{figure}[htb!]
\centering
\includegraphics[scale=0.5]{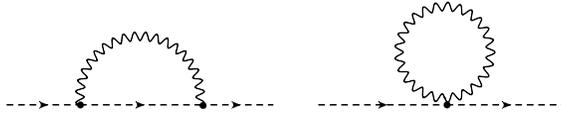}
\caption{One-loop Feynman diagrams contributing to the renormalization of the ghost field.  
A wavy (dotted) line represents gluons (ghosts).}
\label{figZc}
\end{figure}

The ghost field renormalization enters the evaluation of $Z_g$ (see subsection \ref{app:Z}\ref{B4}) and it can be extracted from the RI$'$ condition
\be
\lim_{a\rightarrow 0} \left
    [Z_c^{L,\rm RI'}(a \bar{\mu})\frac{\Sigma^L_c(q,a )}{q^2}\right ]_{q^2=\bar{\mu}^2}=1,
\ee
where $\Sigma^L_c(q,a )$ is the ghost self energy up to one-loop computed from the diagrams in Fig.~\ref{figZc}, namely
\be
\Sigma^L_c(q,a)=q^2+{\cal O}(g^2).
\ee

The generic form of $Z_c^{L,\rm RI'}$ is
\be
Z_c^{L,\rm RI'} = 1 + \frac{g^2 N_c}{16\pi^2}
\Bigl[e_c - 1.2029 \alpha -\frac{1}{4}\left( 3 - \alpha \right)
\log\left(a^2\,\bar{\mu}^2\right) \Bigr].
\ee
The numerical values of the coefficient $e_c$ are listed in Table~\ref{tb:eCoef} for all gluon actions we have considered.

\subsection{Renormalization of Fermion Field ($Z_\psi$) and Mass ($Z_m$)}

In order to obtain the  renormalization functions of fermionic operators we also compute the quark field renormalization, $Z_{\psi}$, as a prerequisite. 

\begin{figure}[!ht]
\centering
\includegraphics[scale=0.5]{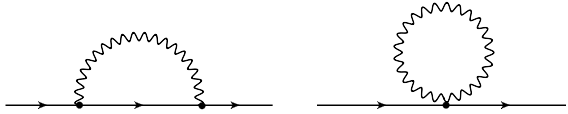}
\caption{One-loop Feynman diagrams contributing to the renormalization of the fermion field.  
A wavy (solid) line represents gluons (fermions).}
\label{figZq}
\end{figure}

$Z_{\psi}$ is extracted from the RI$'$ renormalization condition on the fermion self energy $\Sigma^L_{\psi}(q,a )=i\slashed{q}+ m +{\cal O}(g^2)$, namely
\be
\lim_{a \rightarrow 0} \left[Z_{\psi}^{L,\rm RI'}(a \bar{\mu})\,
{\rm tr}\left(\Sigma_{\psi}^L(q,a )\,\slashed{q}\right)
/(4i\,q^2)\right]_{q^2=\bar{\mu}^2}=1.
\label{ZpsiRule}
\ee
The trace here is over Dirac indices; a Kronecker delta in color and in flavor indices has been factored out of the definition of
$\Sigma_{\psi}^L$. 
The Feynman diagrams contributing to $\Sigma_{\psi}^L$ are shown in Fig.~\ref{figZq}.
Our result for $Z_{\psi}$ is
\be
Z_\psi^{L,\rm RI'} = 1 + \frac{g^2 C_F}{16
  \pi^2}\Bigl[e_\psi - 4.7920 \alpha + \alpha\log\left(a^2\,\bar{\mu}^2\right) \Bigr].
\ee

The part of $\Sigma_{\psi}^L$ proportional to the unit matrix in Dirac space leads to the value of $Z_m$.
Our result for $Z_{m}$ is
\be
Z_m^{L,\rm RI'} = 1 + \frac{g^2 C_F}{16
  \pi^2}\Bigl[e_m + \alpha - 3\,\log\left(a^2\,\bar{\mu}^2\right) \Bigr].
\ee
The numerical values of the coefficients $e_\psi$ and $e_m$ are listed in Table~\ref{tb:eCoef}.

\subsection{Gluon Field Renormalization $Z_A$}

\begin{figure}[htb!]
\centering
\includegraphics[scale=0.7]{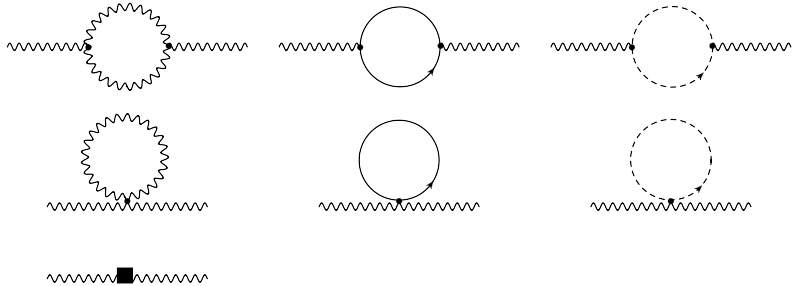}
\caption{One-loop Feynman diagrams contributing to the renormalization of the gluon field.  
A wavy (solid, dotted) line represents gluons (fermions, ghosts). A solid box denotes a vertex from the measure part of the lattice action.}
\label{figZA}
\end{figure}

The renormalization for the gluon field, $Z_A$, can be evaluated from the gluon propagator $G^L_{\mu\,\nu}(q,a )$ with radiative corrections, namely
\begin{equation}
G^L_{\mu\,\nu}(q,a )=\frac{1}{q^2}\left [\frac{\delta_{\mu\,\nu}-q_{\mu}q_{\nu}/q^2}{\Pi_T(a q)}
+\alpha\,\frac{q_{\mu}q_{\nu}/q^2}{\Pi_L(a q)}\right],
\label{GluonProp}
\end{equation}
where the one-loop contributions to the transverse ($\Pi_{T}$) and longitudinal ($\Pi_{L}$) parts of the gluon self-energy, $\Pi_{T,L}(a q)=1+{\cal O}(g^2)$ are obtained from the diagrams of Fig.~\ref{figZA}. 
The normalization condition is
\begin{eqnarray}
\lim_{a \rightarrow 0} \left [ Z_A^{L,\rm RI'}(a \bar{\mu})\, \Pi_T(a q) \right ]_{q^2=\bar{\mu}^2} &=& 1 .
\label{ZAZaRules}
\end{eqnarray}

Our result up to one-loop is
\be
Z_A^{L,\rm RI'} = 1 + \frac{g^2}{16
  \pi^2}\Bigl[N_c \left( e_{A,1}- 0.8863 \alpha +
\frac{1}{4}\alpha^2\right) + \frac{1}{N_c}e_{A,2} -2.1685 N_f 
+ \left( \frac{2}{3} N_f- \frac{13}{6}N_c + \frac{1
  }{2} \alpha N_c\right)\log\left(a^2\,\bar{\mu}^2\right)
\Bigr],
\ee
($N_f$ stands for the number of flavors). The numerical values of the coefficients $e_{A,1}$ and $e_{A,2}$ are listed in Table~\ref{tb:eCoef}.

One may similarly deduce the value of $Z_\alpha$ [see Eq.~(\ref{B5})] from the longitudinal part $\Pi_L(aq)$ of the gluon self-energy; as mentioned before, $Z_\alpha$ receives no one-loop contribution.

\subsection{Coupling constant renormalization $Z_g$}
\label{B4}

$Z_g$ can be extracted either from the gluon-quark-quark Green's function, or equivalently from the gluon-ghost-ghost Green's function $G^{L}_{A\bar{c}c}$; we have chosen to compute the latter\footnote{We have checked, via explicit computation in the Feynman gauge, that the two determinations lead to identical results.}. It is customary to renormalize the strong coupling constant in the $\MSbar$ scheme also when RI$'$ schemes are adopted for the operators. The same choice is made here.

\begin{figure}[htb!]
\centering
\includegraphics[scale=0.7]{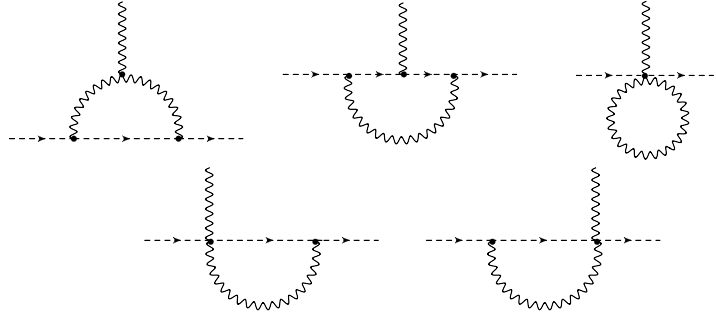}
\caption{One-loop Feynman diagrams contributing to $G^{L}_{A\bar{c}c}$.  
A wavy (dotted) line represents gluons (ghosts).}
\label{figZg}
\end{figure}

The corresponding renormalization condition\footnote{Eq.~(\ref{ZgRules2b}) is evaluated at vanishing ghost momentum; $q$ stands for the ghost/gluon momentum.}
\be
\lim_{a \rightarrow 0} \left[Z_c^{L,\rm RI'}\,(Z_A^{L,\rm RI'})^{1/2}
Z_g^{L,\rm RI'}G^{L}_{A\bar{c}c}(q,a)\right]
_{q^2=\bar{\mu}^2} = G^{{\rm finite}}_{A\bar{c}c},
\label{ZgRules2b}
\ee
where the expression $G^{{\rm finite}}_{A\bar{c}c}$ is required to be the same as the one stemming from the continuum
\be
\lim_{\epsilon\rightarrow 0} \left[Z_c^{DR,\rm RI'}\,(Z_A^{DR,\rm RI'})^{1/2}
Z_g^{DR,\rm RI'}G_{A\bar{c}c}(q)\right]_{q^2=\bar{\mu}^2} = G^{{\rm finite}}_{A\bar{c}c}.
\label{ZgRulesB}
\ee
[In the above equation $Z_g^{DR,\rm RI'}$ is required to eliminate only the pole parts of the left-hand side, without additional finite terms; hence, it is trivially equal to $Z_g^{DR,\MSbar}$.]
Thus, $G^{{\rm finite}}_{A\bar{c}c}$ is found to be
\be
G^{{\rm finite}}_{A\bar{c}c}= 1 + \gtilde\,\Bigl[\Big(\frac{169}{72}  +  \frac{3}{4} \alpha   + \frac{1}{8} \alpha^2  + \frac{1}{2} \alpha \log \left(\frac{\bar{\mu}^2}{q^2} \right)\Big)N_c
- \frac{5}{9} N_f \Bigr].
\ee
The Feynman diagrams contributing to $G^{L}_{A\bar{c}c}$ are shown in Fig.~\ref{figZg}. 
Our result for $Z_g^{L,\rm RI'}$ is
\be
Z_g^{L,\rm RI'} =  1 + \gtilde\,\Bigl[ e_{g,1}\,N_c + \frac{1}{N_c} e_{g,2}
       + 0.5287 N_f + \left(\frac{11}{6} N_c - \frac{1}{3}N_f \right)\log(a^2\,\bar{\mu}^2)\Bigr].
\ee
The numerical values of the coefficients $e_{g,1}$ and $e_{g,2}$ are listed in Table~\ref{tb:eCoef}.

\begin{table}[htb!]
\begin{center}
\begin{tabular}{| c| c| c| c| c| c|}
\hline
\,\,Coefficient\,\, &\,\, Wilson\,\, & \,\,Tree-level Symanzik\,\, & \,\,TILW ($\beta\,c_0=8.30$)\,\, & \,\,Iwasaki\,\, &\,\, DBW2\,\,\\\hline

$e_c$&4.6086&3.7759&3.2208&2.5469&0.9433\\\hline

$e_\psi$&16.6444&13.0233&10.7153&8.1166&2.9154\\\hline

$e_m$&16.9524&13.6067&11.4247&8.8575&2.9060\\\hline

$e_{A,1}$& 22.3157&10.3088&2.4199&-7.2464&-28.5805\\\hline

$e_{A,2}$&-19.7392&-6.6595&2.0039&11.8888&32.2815\\\hline

$e_{g,1}$&-13.4192&-6.5831&-2.0835&3.4235&15.6942\\\hline

$e_{g,2}$&9.8696&3.3297&-1.0019&-5.9444&-16.1407\\\hline
\end{tabular}
\caption{The coefficients $e_c$\,, $e_\psi$\,, $e_m$\,, $e_{A,1}$\,, $e_{A,2}$\,, $e_{g,1}$\,and $e_{g,2}$ for five actions: Wilson, tree-level Symanzik, TILW ($\beta\,c_0=8.30$), Iwasaki and DBW2.}
\label{tb:eCoef}
\end{center}
\end{table}

\subsection{Conversion to the $\MSbar$ scheme}
 
Each renormalization function on the lattice, $Z^{L,\rm RI'}$, may be expressed as a power series in the coupling constant $g$ which is already renormalized in the $\MSbar$ scheme (see Section~\ref{B4}). 

As already mentioned, our one-loop calculations for $Z_c,\,Z_{\psi},\,Z_m,\,Z_A$ and $Z_g$ are performed in a generic gauge with parameter $\alpha^{RI'}$.
The conversion of $\alpha^{RI'}$ to the $\MSbar$ scheme is given by
\be
\alpha^{RI'}=\Bigg(\frac{Z_\alpha^{L,\MSbar}}{Z_\alpha^{L,\rm RI'}}\Bigg)^{-1}\frac{Z_A^{L,\MSbar}}{Z_A^{L,\rm RI'}}\,\,\alpha^{\MSbar}.
\ee
Since $(Z_\alpha^{L,\MSbar}/Z_\alpha^{L,RI'})=(Z_\alpha^{DR,\MSbar}/Z_\alpha^{DR,RI'})=1$ at three loops~\cite{Chetyrkin:2000dq}, it follows
\be
\alpha^{RI'}= \frac{Z_A^{L,\MSbar}}{Z_A^{L,RI'}}\,\alpha^{\MSbar} \equiv \frac{1}{C_A(g^{\MSbar}, \alpha^{\MSbar})} \, \alpha^{\MSbar}\,. 
\label{alphaConversion}
\ee

Since the ratio of $Z$'s appearing in Eq.~(\ref{alphaConversion}) must be {\em regularization independent}, it may be calculated more easily in DR~\cite{Gracey:2003yr}; to one loop, the conversion factor $C_A$ equals
\be
C_A(g,\alpha)=\frac{Z_A^{DR,\rm RI'}}{Z_A^{DR,\MSbar}}=1+\frac{g^2}{36(16\pi^2)}\,\left[\left(9\alpha^2+ 
18\alpha +97\right)\,N_c - 40N_f\right]\,,
\label{CA}
\ee
where hereafter both $g$ and $\alpha$ are expressed in the $\MSbar$ scheme.

Thus, once we have computed the renormalization functions in the $\rm RI'$ scheme we can construct their $\MSbar$ counterparts using conversion factors which, up to the required perturbative order, are given by

\bea
C_c(g,\alpha)&\equiv&\frac{Z_c^{L,\rm RI'}}{Z_c^{L,\MSbar}}=\frac{Z_c^{DR,\rm RI'}}{Z_c^{DR,\MSbar}}=1+\frac{g^2}{16\pi^2}\,N_c\,, \label{Cc} \\
C_{\psi}(g,\alpha)&\equiv&\frac{Z_{\psi}^{L,\rm RI'}}{Z_{\psi}^{L,\MSbar}}=\frac{Z_{\psi}^{DR,\rm RI'}}{Z_{\psi}^{DR,\MSbar}} = 1 - \frac{g^2}{16\pi^2}\,\,C_F\,\alpha, \label{Cpsi}\\
C_{m}(g,\alpha)&\equiv&\frac{Z_{m}^{L,\rm RI'}}{Z_{m}^{L,\MSbar}}=\frac{Z_{m}^{DR,\rm RI'}}{Z_{m}^{DR,\MSbar}} = 1 + \frac{g^2}{16\pi^2}\,\,C_F\,(4 + \alpha). \label{Cmass}
\eea

\end{document}